\documentclass[5p,times]{elsarticle}

\usepackage{hyperref}
\usepackage[caption = false]{subfig}
\usepackage{microtype}
\usepackage{multirow}
\usepackage{graphicx}
\usepackage{epstopdf}
\usepackage{amsmath}
\usepackage[binary-units]{siunitx}
\usepackage{xcolor}
\usepackage{listings}
\usepackage{multicol}

\definecolor{dkgreen}{rgb}{0,0.6,0}
\definecolor{gray}{rgb}{0.5,0.5,0.5}
\definecolor{mauve}{rgb}{0.58,0,0.82}

\usepackage{parcolumns}
\usepackage[autostyle]{csquotes}

\DeclareSIUnit\mykilobyte{KB}

\begin{document}

\begin{frontmatter}




\title{Collectives in hybrid MPI+MPI code: design, practice and performance}

\author{Huan Zhou\corref{cor1}}
 \ead{huan.zhou@hlrs.de}
\author{Jos\'e Gracia}
 \ead{gracia@hlrs.de}
\author{Naweiluo Zhou}
 \ead{naweiluo.zhou@hlrs.de}
\author{Ralf Schneider}
 \ead{schneider@hlrs.de}
\cortext[cor1]{Corresponding author}
\address{High Performance Computing Center Stuttgart (HLRS), University of Stuttgart, 70569 Stuttgart, Germany}




\begin{abstract}

The use of hybrid scheme combining the message passing programming models for
inter-node parallelism and the shared memory programming models for
node-level parallelism is widely spread.
Existing extensive practices on hybrid Message Passing Interface (MPI) plus Open Multi-Processing (OpenMP)
programming account for its popularity.
Nevertheless, strong programming efforts are required to gain
performance benefits from the MPI+OpenMP code. 
An emerging hybrid method that combines MPI and the MPI
shared memory model (MPI+MPI) is promising. However, writing
an efficient hybrid MPI+MPI program -- especially 
when the collective communication operations are involved -- is not to be taken for granted.

In this paper, we propose a new design method to implement
hybrid MPI+MPI context-based collective communication operations.
Our method avoids on-node memory replications 
(on-node communication overheads) that are required
by semantics in pure MPI.
We also offer wrapper primitives hiding all the design details from users,
which comes with practices on 
how to structure hybrid MPI+MPI code with these primitives.
Further, the on-node synchronization scheme required by our method/collectives gets optimized.
The micro-benchmarks show that our collectives are
comparable or superior to those in pure MPI context.
We have further validated the effectiveness of the hybrid MPI+MPI model 
(which uses our wrapper primitives) 
in three computational kernels, by comparison to the
pure MPI and hybrid MPI+OpenMP models.

\end{abstract}

\begin{keyword}
MPI, MPI shared memory model, collective communication, hybrid programming

\end{keyword}

\end{frontmatter}


\section{Introduction}
\label{sec:intro}

For decades the Message Passing Interface (MPI)~\cite{MPISpec} has been a dominant parallel programming model 
in the area of high-performance computing (HPC).
It is widely utilized by applications of interest to various fields and will
continue to be prosperous for its efficiency, adaptivity, and portability.
Nowadays, the computational capability of a single processor grows in a way that increases
its number of computational cores, which strengthens the 
hierarchical memory structure (shared memory within nodes and message passing across nodes).
Memory technology, however, lags behind processor technology.
This dilutes per-core-memory in the current commodity supercomputers.
Traditionally, the applications that are written in pure MPI may face two 
problems. One is the latency, where extra memory copings
are internally required by MPI semantics,
and the other is the memory utilization, where some (on-node) copies of replicated data 
are needed when memory is partitioned across multiple cores for separate address space.
In this regard, the reduced per-core-memory is abused.
Partitioned Global Address Space (PGAS) and
hybrid programming models could be the solutions to
the above two problems.
PGAS, such as Unified Parallel C (UPC)~\cite{upc} and OpenSHMEM~\cite{openshmem},
provides convenient
access to shared global address space.
However, migrating existing MPI parallel programs to 
another PGAS model will burden the users with
a large amount of rewriting work.
Conversely, the hybrid model offers an incremental
pathway to extend existing MPI programs
by combining MPI (inter-node parallelism)
and a shared memory programming approach (node-level parallelism).
Open Multi-Processing (OpenMP)~\cite{OpenMP} is the most \mbox{frequently-used}
shared memory programming model~\cite{bernholdt2017survey}. 
The simplest approach
is to incrementally add OpenMP directives
to the computationally-intensive parts of the existing MPI code,
which is also called OpenMP fine-grained parallelism~\cite{cappello2000mpi}.
This approach can produce serial sections that are only executed 
by the master thread.
Coupled with the extra overheads
from shared memory threading, such hybrid implementation may hardly outperform 
the pure MPI implementation when the scaling of the MPI implementation is still good
\cite{techrep/hybridmpi+openmpi,krawezik2006performance}.
When the scalability of the pure MPI code suffers
a lot, the hybrid one could perform better with less communication time
\cite{conf/pdp/RabenseifnerHJ09,conf/ipps/DrosinosK04}.
There are still two new hybrid parallel programming methods:
MPI+UPC~\cite{dinan2010hybrid} and MPI+OpenSHMEM~\cite{jose2012supporting}.
They attract little attention, since a profound 
grasp of both MPI and OpenSHMEM or UPC is needed to 
write efficient applications.

Further, an innovative hybrid programming approach combining MPI and the
MPI Shared Memory (SHM) model emerges (MPI+MPI~\cite{hoefler2013mpi+}).
The MPI SHM model
\cite{Introduction:SHM,conf/europar/ZhouIG15,conf/pvm/HoeflerDBBBBGKT12,karlbom2016performance}
is process-based and
introduced in the \mbox{MPI-3} standard for
supporting shared memory address space
among MPI processes on the same node. 
In the hybrid MPI+MPI model,
the on-node shared data is logically partitioned and 
a portion of it is affinity to each process.
Compared with the MPI model, the computational parallelism in the MPI+MPI version
stays unchanged and the on-node communication overhead
is eliminated. Therefore, this hybrid scheme
is expected to benefit performance, even when the pure
MPI applications are already good in scalability.
Nevertheless, there are so far very limited practices to guide the
users in writing scalable as well as efficient hybrid MPI+MPI applications,
except the study~\cite{hoefler2013mpi+} that demonstrates
a hybrid MPI+MPI programming paradigm featuring the point-to-point
communication operations (e.g., halo exchanges).
However, this paradigm does not strictly follow the shared memory
programming scheme that demands only one (shared) copy of replicated data
among on-node processes.
Besides the point-to-point communication operations, MPI provides a rich
suite of collective operations that involve a group
of processes rather than a pair of processes.
The MPI collectives are important, as they are frequently invoked
in a spectrum of scientific applications or kernels~\cite{NASA}.
They always appear in performance-critical sections of these applications.
If the hybrid MPI+MPI code continues to harness the standard MPI collectives as
the pure MPI code does, scalable performance is difficult to achieve.
Therefore, designing hybrid MPI+MPI context-based 
collective communication operations and 
creating experience in writing scalable and efficient
hybrid MPI+MPI programs including these collectives are inspired.

In the pure MPI version, the collectives give a copy of the 
result to every on-node process, which is dispensable in
the hybrid MPI+MPI version when each process proceeds to 
read the result with \textit{visible} or no change to it. 
This is the case in most of the existing
applications or kernels containing the collective operations~\cite{NASA}
and the benchmarks used in this paper as well.
The \textit{visible} change, as the name implies, the change is visible
to other processes (shared between processes). 
Further, the \textit{visible} changes to the same data can be
synchronized by using the method proposed in~\cite{zhou2014dart}.
Conversely, \textit{invisible} change signifies private change, which 
entails a copy of the accessed data.
Previously, we discussed the programmatic
differences between the approach of 
collectives in the hybrid MPI+MPI context with the standard one
in the pure MPI context~\cite{zhou2019mpi}.
In this paper, we (take \textit{allgather}, broadcast and \textit{allreduce} as concrete cases)
further explore the challenges associated with
designing the hybrid MPI+MPI context-based collective operations
and writing an efficient hybrid MPI+MPI code with an acceptable 
number of lines.
The main contributions of our work are fourfold:
\begin{enumerate}
\item Besides \textit{allgather} and broadcast, we describe the design method
of \textit{allreduce} in the hybrid MPI+MPI context.
We provide the users with fully-functional MPI wrappers
that hide all the design details of our
collectives and demonstrate the necessity of applying
these wrappers to the hybrid MPI+MPI programming. 
\item We highlight all synchronization points that are inherently required by our 
\textit{allgather}, broadcast and \textit{allreduce} to guarantee data integrity 
inside nodes. We then discuss
how to implement them with minimal overhead.
\item We perform a series of micro-benchmarks to 
first quantify
the implementation overhead brought by our collectives and then 
compare our  
collectives and the standard MPI collectives,
under the same distribution of workload on all cores.

\item We conduct three case studies 
to show the benefit of the hybrid MPI+MPI code
calling our collectives
over the pure MPI and hybrid MPI+OpenMP code.
\end{enumerate}

The paper is organized as follows.
In the next section, we briefly give the related work.
Section~\ref{sec:background} describes 
the MPI SHM model that forms 
a basis for our hybrid MPI+MPI context-based collectives,
and the skeleton of a simple hybrid MPI+MPI program.
Section~\ref{sec:impl} starts with a description of 
our collectives, provides the users with wrapper functions, presents examples written in the
hybrid MPI+MPI context and proposes a relatively light-weight
synchronization method.
In Section~\ref{sec:eval}, the experimental results and analyses
based on the micro- and kernel-level benchmarks are demonstrated.
Section~\ref{sec:conclusion} discusses and concludes our paper.

\section{Related work}
\label{sec:related-work}

In the early stage of optimizing MPI collective operations, the researchers 
put much effort into studying optimal algorithms. This leads to the coexistence
between different algorithms catering to different message sizes and
numbers of processes\cite{thakur2005optimization, pjevsivac2007performance}.
The  high-performance
implementations of MPI, such as MPICH~\cite{MPICH}, 
Intel MPI~\cite{IntelMPI} and Open MPI~\cite{OpenMPI}, thus choose the most appropriate algorithm to use at runtime.

The prevalence of clusters of shared memory nodes
highlights a hybrid architecture combining distributed
(across nodes) and shared memory (constrained within a single node).
The optimized works have shifted to 
distinguish between intra-node and inter-node communication
(aka., hierarchical algorithm
\cite{hasanov2015hierarchical, zhu2009hierarchical, traff2014mpi}).
The hierarchical algorithm has been adopted by
the existing MPI implementations and pays off.
MPI collectives are expected to be highly tuned for
shared memory as well as distributed architecture.
In~\cite{conf/pvm/GrahamS08, conf/ccgrid/MamidalaKDP08}, 
the MPI collectives are optimized by using the shared cache
as an intra-node data transfer layer.
Nowadays, a typical shared memory node features non-uniform memory access (NUMA) architecture,
the NUMA-aware shared memory MPI collectives are thus proposed to
further minimize the inter-NUMA (intersocket) memory traffic~\cite{conf/hpdc/LiHS13}.
Besides that, remote direct memory access (RDMA)
is used for inter-node communication to improve performance~\cite{mamidala2006efficient, qian2007rdma}.
All the aforementioned optimizations pave the way to the maturity of MPI collectives.
Furthermore,
the scalability of MPI+OpenMP hybrid application
has been improved by making full use of 
idle OpenMP threads to 
parallelize the MPI collectives~\cite{conf/pvm/MaheoCPJ14}.

\section{MPI+OpenMP versus MPI+MPI}
\label{sec:background}

In this section we describe the MPI-3 shared memory model
and two-level of communicator splitting. They are foundations
of the hybrid MPI+MPI programming model.
We further provide a brief comparison between two skeleton programs of
MPI+OpenMP and MPI+MPI containing collective
communication operations.

\subsection{MPI+OpenMP}
\label{sec:back:MPIOpenMP}

\begin{figure}[tbp]
 \begin{center}
  \includegraphics[width=0.5\textwidth,height=0.25\textheight]{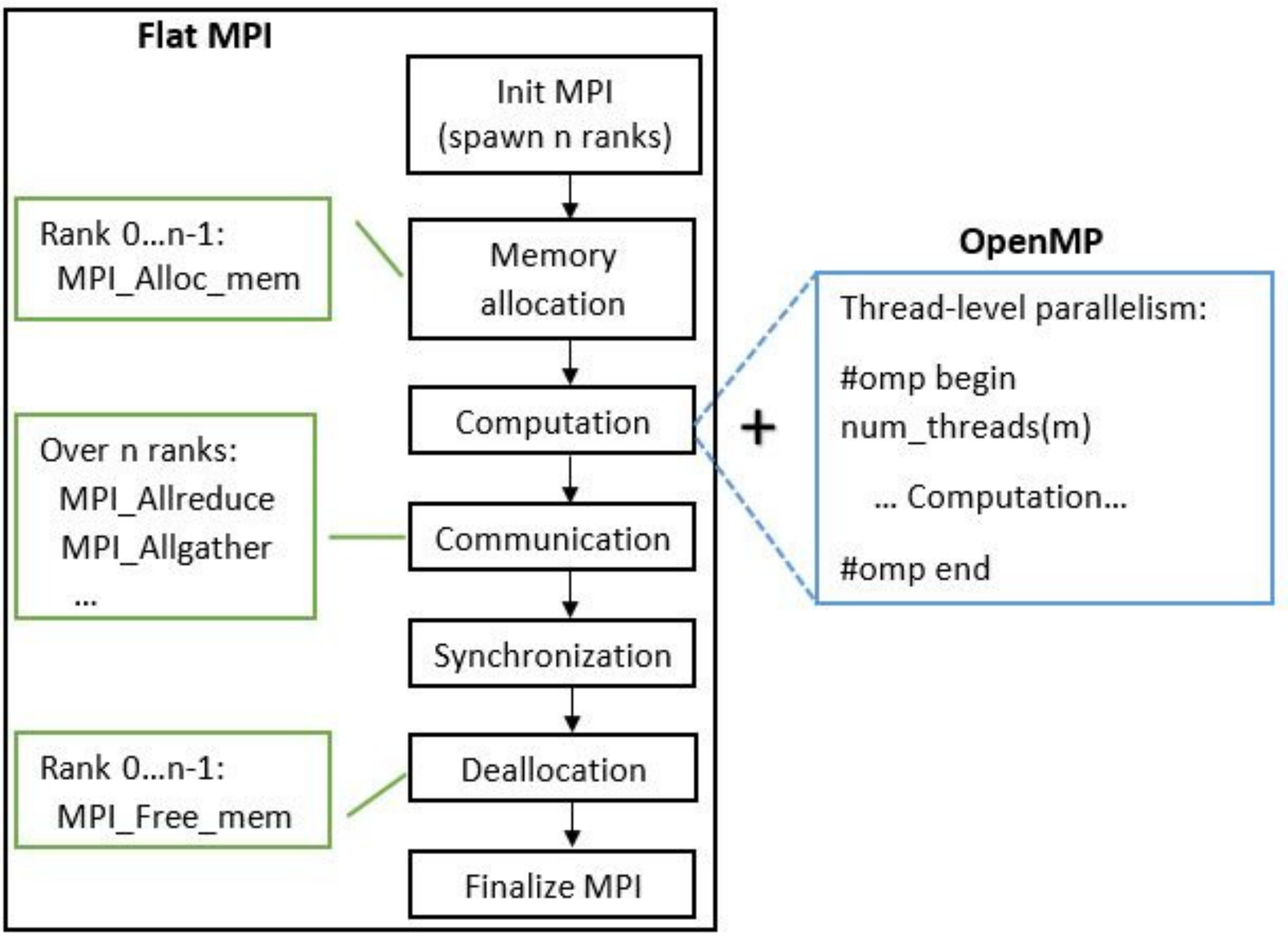}
 \end{center}
 \caption{The workflow of the hybrid MPI+OpenMP programming model with the fine-grained on-node parallelization approach}
 \label{fig:workflow:MPIOpenMP}
\end{figure}

In the hybrid MPI+OpenMP hybrid model, MPI is used for communication across
distributed memory nodes and OpenMP is responsible for on-node parallelization.
Assuming there is a cluster of $n$ shared memory nodes, each of which consists of $m$ computational cores.
Figure~\ref{fig:workflow:MPIOpenMP} illustrates the workflow of
writing a hybrid MPI+OpenMP program (run on the $n$ nodes)
by using the OpenMP fine-grained parallelization approach~\cite{cappello2000mpi}.
At the period of initialization, $n$ MPI processes are spawned and each
of them is allocated on distinct node.
The left part shows that each MPI process allocates or frees memory 
which has its own address space (not addressable by each other).
The computation component
will resort to OpenMP directives, which is demonstrated
in the right part of the figure.
Here, each process spawns $m$ threads (each thread is pinned to a core)
executing the computation concurrently.
In this scenario, the standard MPI collective
communication operations over the 
$n$ MPI processes are directly harnessed.
The advantage of OpenMP offering incremental approach towards parallelization
facilitates the porting work from MPI code to hybrid MPI+OpenMP one.
In the hybrid MPI+OpenMP version, however, creating the same parallelism as in
the MPI version is daunting and needs plenty of human efforts,
which will, in turn, reduce the advantage in using OpenMP.
Besides the fine-grained parallelism, another parallelism approach
of coarse-grained is also studied but not so mature as the former.
Therefore, the hybrid MPI+OpenMP program with
fine-grained parallelism is considered as one of the baselines
for evaluating the hybrid MPI+MPI program in Section~\ref{sec:eval-app}.

\subsection{MPI+MPI}
MPI-3 extends the standard MPI with the shared memory programming
that supports direct load/store operations on a single node.
In this section, we introduce the shared memory and 
bridge communicators.
The concept of the shared memory window is also critical for us
to understand how MPI SHM exposes a global view of on-node memory to the users.
With all these knowledge in hand, we introduce the workflow of
writing a typical hybrid MPI+MPI program running on the aforementioned
cluster of $n$ nodes.

\subsubsection{Two level of communicator splitting}

The function {\em MPI\_Comm\_split\_type} is called with 
the parameter of {\em MPI\_COMM\_TYPE\_SHARED} to
divide the communicator into discrete node-level communicators.
Each node-level (aka. shared memory)
communicator identifies a group of processes
that are connected to the same shared memory system, inside
which all processes can perform load/store operations instead of 
explicit remote memory access (RMA).
Besides the shared memory communicator,
the hybrid MPI+MPI programming model entails
an across-node communicator to serve for the explicit communication
between processes residing on different nodes.
A process per node (often with the lowest rank) is chosen 
as a $leader$ to take responsibility for the 
data exchanges across nodes, while the other on-node processes
are viewed as its $children$.
The across-node communicator acts as a bridge between nodes and
thus is also called bridge communicator~\cite{traff2014mpi},
which is formed by calling {\em MPI\_Comm\_split}.

\subsubsection{MPI shared memory window}
\label{sec:background:sharedmem}

The usage of {\em MPI\_Win\_allocate\_shared} is crucial to
create a window spanning a region of addressable shared memory 
with an individual size that is contributed by
each on-node process.
By default, the memory in a window is allocated contiguously in hardware.
Creating non-contiguous memory is also possible when the 
parameter of {\em alloc\_shared\_noncontig} is set to true.
The function {\em MPI\_Win\_shared\_query} is used
to obtain the base pointer to the beginning of the shared memory
segment contributed by another process.
This base pointer allows the allocated memory to be accessed 
with immediate \mbox{load/store} instructions by all on-node processes.
Intuitively, the function {\em MPI\_Win\_sync} is defined to synchronize
between the private and public window copies. Nowadays,
the majority of hardware architectures feature a
unified memory model, where
the public and private copies can be maintained consistent implicitly.
Nevertheless, the usage of {\em MPI\_Win\_sync} is still valuable in achieving
a memory synchronization when there are concurrent accesses to the same
memory location by different on-node processes.

\subsubsection{Workflow}
\label{sec:background:sharedmemworkflow}
\begin{figure}[tbp]
 \begin{center}
  \includegraphics[width=0.42\textwidth,height=0.31\textheight]{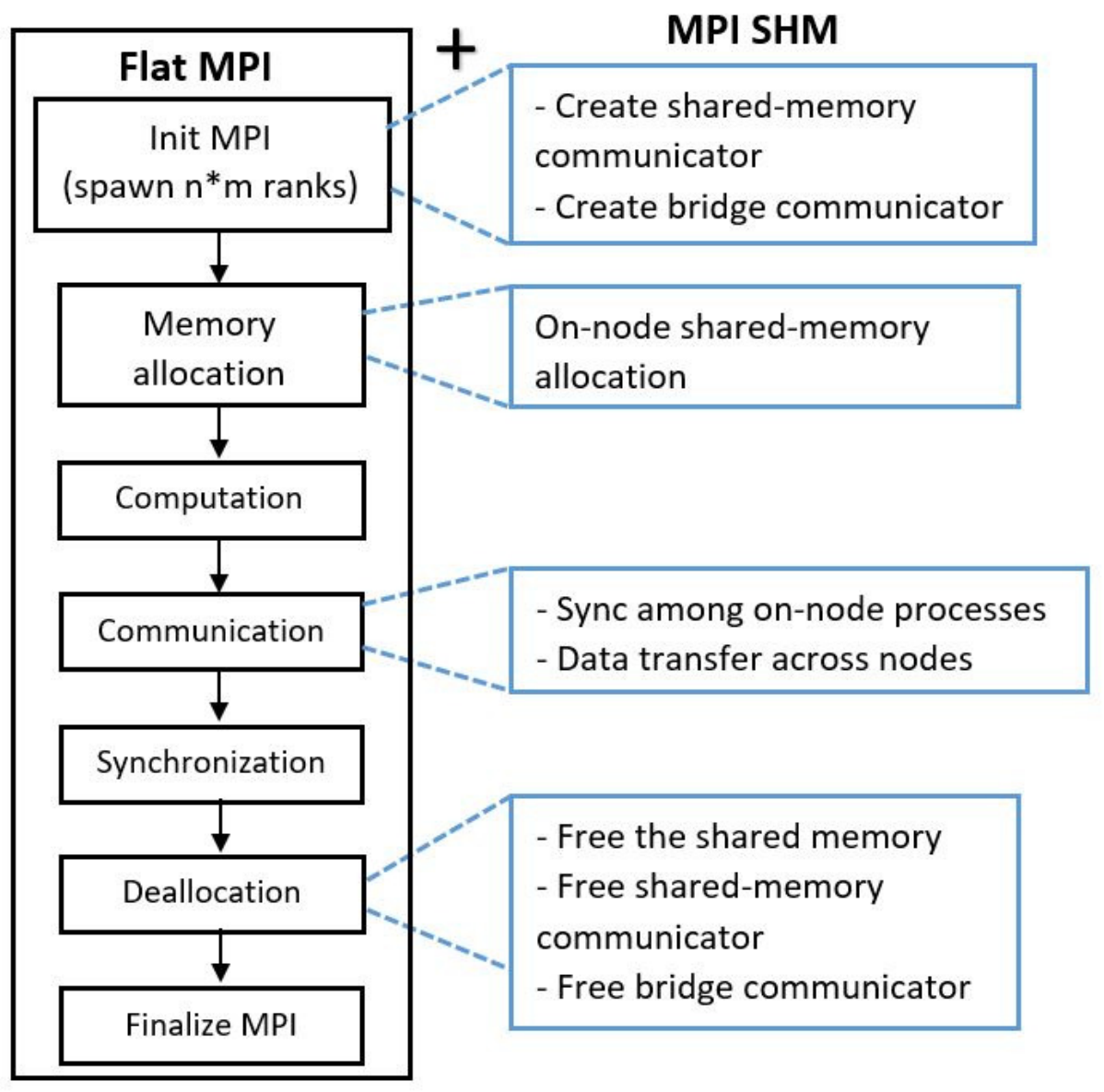}
 \end{center}
 \caption{The workflow of the hybrid MPI+MPI programming model}
 \label{fig:workflow:MPIMPI}
\end{figure}

Theoretically, the migration from pure MPI programs to hybrid MPI+MPI ones
should be smooth due to their interoperability.
Figure~\ref{fig:workflow:MPIMPI} presents the hybrid MPI+MPI programming pattern,
where the right part presents the possible rewriting efforts for achieving this hybridization. Here,
$n*m$ (equal to the number of available cores) MPI processes are spawned during initialization,
where the shared memory and 
bridge communicators are required to be generated.
The $leader$ allocates the entire shared memory region
for all on-node processes and then its $children$ attach
to a separate portion of this shared memory region.
When a global communication operation happens,
the shared region can be accessed by executing \mbox{load/store} instructions,
with all the node-level synchronizations to guarantee
its consistent status among on-node processes.
This shared region can also certainly be touched 
by the processes residing on different nodes via RMA, collective
and point-to-point communication operations.
Before finalizing the program,
the above-mentioned two communicators and the shared region should be deallocated.

Our comparison between hybrid MPI+MPI and pure MPI or hybrid MPI+OpenMP
consists in two aspects: programmability and performance. 
In detail,
preparing the aforementioned communicators and shared memory windows
for a hybrid MPI+MPI program can be tedious.
Moreover, a hybrid MPI+MPI program is error-prone when 
the users mishandle the node-level synchronization operations.
The rewriting efforts are clearly not negligible.
Therefore, wrapper functions encapsulating these rewriting
details should be available to the users for well-modularized programming.
Like MPI, the problems/tasks are also forced to be decomposed and evenly
assigned to separate processes for locality in hybrid MPI+MPI.
Hence unequal parallelism will not be the reason for
the performance benefits of MPI over hybrid MPI+MPI.
Besides measuring the performance of 
our collectives, 
their implementation overheads
need to be considered when the holistic performance of a hybrid MPI+MPI program is assessed.

\section{Implementation and practices}
\label{sec:impl}

In this section, we present several generic wrapper
interfaces that should always be included to enable a 
hybrid MPI+MPI program with the pattern shown in Figure~\ref{fig:workflow:MPIMPI}.
We take three typical collectives
({\em MPI\_Allgather}, {\em MPI\_Allreduce},
and {\em MPI\_Bcast})
for example, to describe our efforts in implementing
the hybrid MPI+MPI context-based collectives by assuming that
the block-style rank placement is employed.
I.e. the
consecutive ranks fill up each compute (shared memory) node before moving to the next.
Each of the standard MPI collective communication interfaces referenced above
has a counterpart in our hybrid approach.
The counterparts change the parameters slightly.
In addition, there could be specific wrapper functions
contributing to their implementations.
Based on the wrapper primitives, we give a practice in building
a prototypical hybrid MPI+MPI code, where the \textit{allgather} is involved. Furthermore,
the link\footnote{https://github.com/HyMPIMPIColl/BenchHyCollWithWrapper}
provides examples describing the usage of our broadcast and \textit{allreduce} in the hybrid MPI+MPI context.
We prove that these wrapper interfaces play an important role in
improving the productivity of the hybrid MPI+MPI application developers
by unveiling the implementation details hidden in them.
According to Figure~\ref{fig:workflow:MPIMPI},
on-node synchronization should be carefully considered
for the hybrid approach inside the communication component.
We thus discuss how the node-level synchronizations could be
implemented for benefiting the performance of the
hybrid MPI+MPI programs.

\subsection{Common wrapper primitives}
In all MPI+MPI programs embracing the collective operations, the 
steps manipulating communicators and shared regions are common places.
To obviate code duplication,
we wrap all the common steps into the corresponding wrapper functions,
whose interfaces are demonstrated in Figure~\ref{fig:impl:generic}.
The structure {\em comm\_package} defines variables associated with
the shared memory and bridge sub-communicators.
The function \mbox{\em Wrapper\_MPI\_ShmemBridgeComm\_create} takes
a communicator as input parameter and
returns an instance of the above structure.
Aside from the {\em MPI\_COMM\_WORLD}, 
other communicators deriving from it are supported by this function
for complex use cases.
The $msize$, $bsize$ and $flag$
-- in function {\em Wrapper\_MPI\_Sharedmemory\_alloc} -- 
are parameters defined to determine
the total size (in bytes) of a shared region allocated in the $leader$.
These two wrapper functions are both one-off activities
whose overheads are evaluated in Section~\ref{sec:eval:overhead}.
The function
{\em Wrapper\_Get\_localpointer} needs to be invoked to
output a local pointer pointing to the shared memory location with
affinity to the calling process. 
In the end, we need to explicitly
deallocate the communicators via the wrapper function \mbox{\em Wrapper\_Comm\_free}.

\begin{figure}[tbp]
\begin{lstlisting}[mathescape, escapechar=']
/* The structure of data type comm_package */  					
struct comm_package
{
		MPI_Comm		shmem_comm;
		MPI_Comm		bridge_comm;
		int			shmemcomm_size;//Size of shared memory communicator
		int			bridgecomm_size;//Size of bridge communicator
};
/* Two level of communicator splitting */
void Wrapper_MPI_ShmemBridgeComm_create(MPI_Comm par_comm,
					struct comm_package *comm_handle);
/* Shared memory allocation */
void Wrapper_MPI_Sharedmemory_alloc(int msize, int bsize, 
  			int flag, struct comm_package *comm_handle,
  			void **shmem_addr, MPI_Win *winPtr);
/* Affinity */
void Wrapper_Get_localpointer(void *start_addr,
					int rank, int dsize, void **local_addr); 
/* Free shared memory and bridge communicators */
Wrapper_Comm_free(struct comm_package *comm_handle);
\end{lstlisting}
\caption{The wrapper interfaces handling with communicators and regions of shared memory.}
\label{fig:impl:generic}
\end{figure}

\subsection{Allgather}
\label{sec:impl-allgather}

\begin{figure}[h!]
\centering
 \subfloat[\textit{Allgather} in the pure MPI version]{\label{fig:allgather:nativ}\includegraphics[width=0.38\textwidth,height=0.15\textheight]{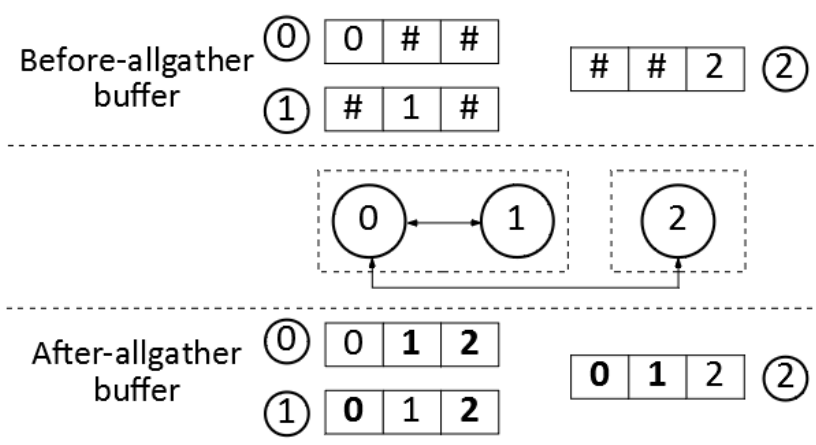}}\\
 \subfloat[\textit{Allgather} in the hybrid MPI+MPI version]{\label{fig:allgather:hy}\includegraphics[width=0.32\textwidth,height=0.16\textheight]{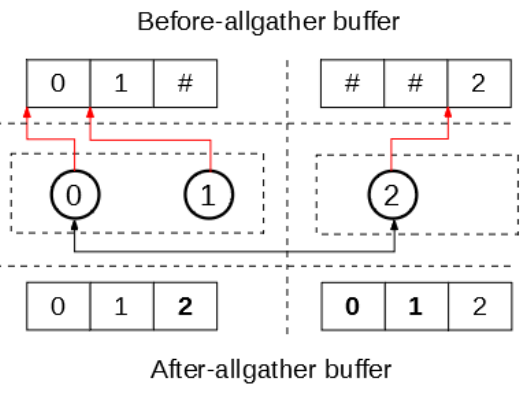}}
\caption{Comparison of the pure MPI context-based and 
hybrid MPI+MPI context-based \textit{allgather} according to the changes in
buffers for each process.
$\sharp$: empty element; bold font: gathered element from other processes;
black arrow: an inter-process communication;
red arrow: a local pointer.}
\label{fig:allgathernativehy}
\end{figure}

This section first describes the implementation
dissimilarities of the pure MPI context-based \textit{allgather} 
({\em MPI\_Allgather})
and the hybrid MPI+MPI context-based \textit{allgather} in Figure~\ref{fig:allgathernativehy}.
The latter is called \mbox{\em Wrapper\_Hy\_Allgather} in our
hybrid MPI+MPI version.
Our previous paper~\cite{zhou2019mpi} can be referred to for a more elaborate description.
Then we focus on the practices in writing hybrid MPI+MPI code based
on our wrapper interfaces and compare it with the one without the use of them.

Both legends in Figure~\ref{fig:allgathernativehy} describe their implementation approaches 
according to the usage and status of buffer in each process
instead of the switch between the \textit{allgather} algorithms (e.g., recursive doubling or ring).
Shown in Figure~\ref{fig:allgather:nativ},
initially process i assigns a valid value to the i-th
element as its local data, that is about to
be sent to other processes. 
After this operation, the message sent from each process is placed in rank order
in all processes' \mbox{after-allgather} buffers,
where the copies of replicated data inside node are noticed.
Besides, the intra-node communication involves extra buffer
allocation and copies, which are
determined by the underlying MPI library and occur transparently to the user.
Unlike the \textit{allgather} in the pure MPI version, 
only one copy of buffer, which is allocated as a shared memory segment,
is demanded on a node in our \textit{allgather}, shown in Figure~\ref{fig:allgather:hy}.
This buffer is shared among all the on-node processes and thus
the intra-node communication is eliminated.
Therefore, in our \textit{allgather} the $leaders$ (comprise process $0$ and $2$), as the
representatives of the two nodes, are required to exchange
all the valid messages.
The irregular \textit{allgather} variant ({\em MPI\_Allgatherv})
is leveraged for this across-node data exchanges, 
due to that the valid message size could vary from one node to another.
In order to achieve the same
computational parallelism as the standard \textit{allgather},
the on-node shared region is evenly partitioned into
separate portions, each of which builds an affinity with a
process via a pointer.
This is done before the \mbox{\em Wrapper\_Hy\_Allgather} is executed.

\lstset{
  language=C++,
  showstringspaces=false,
  columns=flexible,
  frame=tb,
  xleftmargin=2em,
  framexleftmargin=2em,
  basicstyle={ \footnotesize\ttfamily},
  numbers=left,
  numberstyle=\tiny,
  keywordstyle=\color{blue},
  commentstyle=\color{dkgreen},
  stringstyle=\color{mauve},
  breaklines=true,
  breakatwhitespace=true,
  tabsize=1, 
  emphstyle=\underline,
}
\begin{figure}[h!]
\begin{lstlisting}[mathescape, escapechar=']
struct comm_package comm_handle;
struct allgather_param param_handle;
MPI_Win win;
double *result_addr, *s_buf, *r_buf;
s_buf = r_buf = NULL;
int    nprocs, *sharedmem_sizeset, rank;
Wrapper_MPI_ShmemBridgeComm_create(MPI_COMM_WORLD, 
					&comm_handle);
MPI_Comm_size(MPI_COMM_WORLD, &nprocs);
MPI_Comm_rank(MPI_COMM_WORLD, &rank);
Wrapper_MPI_Sharedmemory_alloc(msg, sizeof(double),
		 		nprocs, &comm_handle, (void**)&r_buf, &win);
Wrapper_ShmemcommSizeset_gather(&comm_handle, 
					&sharedmem_sizeset);
Wrapper_Create_Allgather_param(msg, &comm_handle, 
					sharedmem_sizeset, &param_handle);
Wrapper_Get_localpointer(r_buf, rank,
					msg*sizeof(double), (void**)&s_buf);
for(int i = 0; i < msg; i++){ s_buf[i] = i; }  
Wrapper_Hy_Allgather<double>(r_buf, s_buf, msg, 
					MPI_DOUBLE, &param_handle, &comm_handle);
MPI_Win_free(&win);//Free the allocated shared memory
Wrapper_Param_free(&comm_handle, &param_handle);
Wrapper_ShmemcommSizeset_free(&comm_handle,
					sharedmem_sizeset);
Wrapper_Comm_free(&comm_handle);
\end{lstlisting}
\caption{A simple hybrid MPI+MPI example including an \textit{allgather} operation.}
\label{hyallgathertestcode}
\end{figure} 

\begin{figure}[h!]
\begin{lstlisting}[mathescape, escapechar=']
/* Hierarchical communicator splitting '\cite{traff2014mpi}' */ 
comm = MPI_COMM_WORLD;
MPI_Comm_split_type(comm, MPI_COMM_TYPE_SHARED,
     0, MPI_INFO_NULL, &shmem_comm);
MPI_Comm_rank(shmem_comm, &shmemcomm_rank);
leader = 0;
MPI_Comm_split(comm, 
     (shmemcomm_rank==leader)?0:MPI_UNDEFINED,0,
     &bridge_comm);
Every process gets shmemcomm_size 'and' bridgecomm_size;           
MPI_Comm_size(comm, &nprocs); 
msgSize = (shmemcomm_rank==leader)?msg*nprocs:0;
MPI_Win_allocate_shared(msgSize, sizeof(double),
     MPI_INFO_NULL, shmem_comm, &r_buf, &win);                   
if (shmemcomm_rank != leader){
     MPI_Win_shared_query(win, leader, &r_buf);}
MPI_Comm_rank(comm, &rank);
if (bridge_comm != MPI_COMM_NULL){
     sharedmem_sizeset = malloc(.);
     recvcounts = malloc(.);displs = malloc(.);
     MPI_Allgather(shmemcomm_size, sharedmem_sizeset, 
        bridge_comm);
     for (int i = 0; i < bridgecomm_size; i++){
         recvcounts = msg*sharedmem_sizeset[i];
         displs[i] = 0;
         for (int j = 0; j < i; j++)
             displs[i] = recvcounts[j];}}
s_buf = r_buf + msg*rank;
for(int i = 0; i < msg; i++){ s_buf[i] = i; }
if (bridgeComm != MPI_COMM_NULL){// Leaders
     MPI_Barrier(sharedmemComm);
     MPI_Allgatherv(s_buf, r_buf, recvcounts, displs, bridgeComm);
     MPI_Barrier(sharedmemComm);}
else{// Children
     MPI_Barrier(sharedmemComm);
     MPI_Barrier(sharedmemComm);}
MPI_Win_free(&win); 
MPI_Comm_free(shmem_comm);
if (bridge_comm != MPI_COMM_NULL){
     MPI_Comm_free(bridge_comm);free(sharedmem_sizeset);
     free(recvcounts);free(displs);}
\end{lstlisting}
\caption{Pseudo-code that illustrates how to implement the above example (see Figure~\ref{hyallgathertestcode}) 
without the wrapper interfaces.}
\label{hybridallgatherverbosecode}
\end{figure} 

\lstset{
  language=C++,
  showstringspaces=false,
  columns=flexible,
  frame=tb,
  xleftmargin=0em,
  framexleftmargin=0em,
  basicstyle={ \footnotesize\ttfamily},
  numbers=none,
  numberstyle=\tiny,
  keywordstyle=\color{blue},
  commentstyle=\color{dkgreen},
  stringstyle=\color{mauve},
  breaklines=true,
  breakatwhitespace=true,
  tabsize=1, 
  emphstyle=\underline,
}

Figure~\ref{hyallgathertestcode} shows a complete and simple example
(micro benchmark) of how to illustrate a hybrid MPI+MPI program
containing an \textit{allgather} operation by using our wrapper interfaces.
Besides the common wrapper functions defined above, there are
several wrapper functions and data structures specifically 
provided for implementing our \textit{allgather} approach.
The data structure {\em allgather\_param} (line $2$)
stores two integer arrays (i.e., \texttt{recvcounts}
and \texttt{displs}) specifying the receive counts and displacements.
These two arrays are required by our template function {\em Wrapper\_Hy\_Allgather},
which is the counterpart to the {\em MPI\_Allgather} used
in pure MPI version
and thus is the object to be measured in Section~\ref{sec:eval-allgather}.
Lines $13$ and $14$ generate an array (i.e., {\em sharedmem\_sizeset}) that collects the sizes
of all the shared memory communicators. 
The function {\em Wrapper\_Create\_Allgather\_param}
receives this array as input data and returns a value to
{\em param\_handle} of type struct {\em allgather\_param} 
(lines $15$ and $16$).
This function computes the sets of
received counts and displacements for irregular \textit{allgather} and
is also a one-off, which could be amortized
in the future by repeatedly invoking 
{\em Wrapper\_Hy\_Allgather} operation.
In the end, the {\em sharedmem\_sizeset} and {\em param\_handle} 
should be properly freed.

Figure~\ref{hybridallgatherverbosecode} is added to expand the wrapper  
functions of relevance to Figure~\ref{hyallgathertestcode}
and it shows how our design is originally realized in the hybrid MPI+MPI context
without our wrapper interfaces.
Due to space limit, Figure~\ref{hybridallgatherverbosecode} skips the declarations of variables.
Obviously, the program listed in Figure~\ref{hybridallgatherverbosecode} (hereafter called \textit{verbose program})
produces more lines of code (LOC) than that 
demonstrated in Figure~\ref{hyallgathertestcode} (hereafter called \textit{wrapper program}).
In order to better grasp the contribution of these wrapper interfaces,
the positional correspondence between the functionalities of the above two programs
is further generated and shown in Table~\ref{tab:correlation}, where
the leftmost column lists the involved functionalities.
On the right columns, the line numbers indicate the position of the given functionality in each program.
We can observe that each functionality corresponds to one or several wrapper interfaces, which
make the \textit{wrapper program} more structured and readable. Conversely,
the \textit{verbose program} is prone to obscurity or even failure due to that it explicitly handles
the details of all the listed functionalities.
In addition, the hybrid MPI+MPI program developers can benefit from this mapping table
that enables them to better apply our wrapper interfaces to their own applications.
In short,
the above study emphasizes the need for the use of the wrapper interfaces with proven benefits --
better productivity and applicability -- to the hybrid MPI+MPI users.


\begin{table}[thbp]
\footnotesize
\begin{center}
\begin{tabular}{|c|c|c|}
\hline
\multirow{2}{*}{Functionality} & \multicolumn{2}{c|}{Lines} \\ \cline{2-3} 
                               & \textit{wrapper program}     & \textit{verbose program}    \\ \hline
Communicator splitting         & 7-8          & 2-10        \\ \hline
Shared memory allocation            & 11-12        & 12-16       \\ \hline
Fill \texttt{recvcounts} and \texttt{displs} & 13-16        & 18-27       \\ \hline
Get local pointer              & 17-18        & 28          \\ \hline
\textit{Allgather}                      & 20-21        & 30-36       \\ \hline
Deallocation                   & 23-26        & 38-41       \\ \hline
\end{tabular}
\end{center}
\caption{Correspondence between \textit{wrapper program} and \textit{verbose program}.}
\label{tab:correlation}
\end{table}

\subsection{Broadcast}
\label{sec:impl-broadcast}
\begin{figure}[h!]
\begin{lstlisting}[mathescape, escapechar=']
void Wrapper_Get_transtable(MPI_Comm p_comm,
					const struct comm_package* comm_handle, 
					int **shmem_transtable, int **bridge_transtable)
template<class myType>
void Wrapper_Hy_Bcast(myType** bcast_addr,
					myType* start_addr, int msize, int* shmem_transtable, 
					int* bridge_transtable, MPI_Datatype data_type,
					int root, struct comm_package* comm_handle);      
\end{lstlisting}
\caption{The specific wrapper interfaces with respect to our broadcast.}
\label{fig:impl:bcast}
\end{figure}
A broadcast operation happens when one MPI process, called $root$, sends
the same message to every other process.
Likewise,  in our
MPI+MPI context-based broadcast approach,
a region of memory is allocated to store
the broadcast data in each $leader$
and can be shared by its $children$.
Only the $root$ is eligible to alter the broadcast data according
to the MPI broadcast semantics.
All processes on the same node independently read the broadcast
data via a local pointer to the beginning of this shared memory location.
Here, performing the across-node broadcast operation (over all the $leaders$)
is straightforward since the size of the broadcast message remains the same
as that of the pure MPI context-based broadcast.

Broadcast operation is rooted and can only be performed when
the $root$'s rank is correctly given.
Every process can be the $root$ in real world.
This confronts us with the challenge of determining
the relative rank of the $root$ in both the shared memory
and bridge \mbox{sub-communicators}.
Two \mbox{absolute-to-relative} rank translation tables --
{\em shmem\_transtable} and {\em bridge\_transtable} -- are thus generated in
function \mbox{\em Wrapper\_Get\_transtable}.
It brings implementation overhead to our broadcast approach.
The function {\em Wrapper\_Hy\_Bcast} receives the above 
two translation tables as input data to perform
our hybrid broadcast operation.
The above two primitives provided for our broadcast are shown in Figure~\ref{fig:impl:bcast}.

\subsection{Allreduce}
\label{sec:impl-allreduce}

\begin{figure}[h!]
\centering
  \includegraphics[width=0.355\textwidth,height=0.155\textheight]{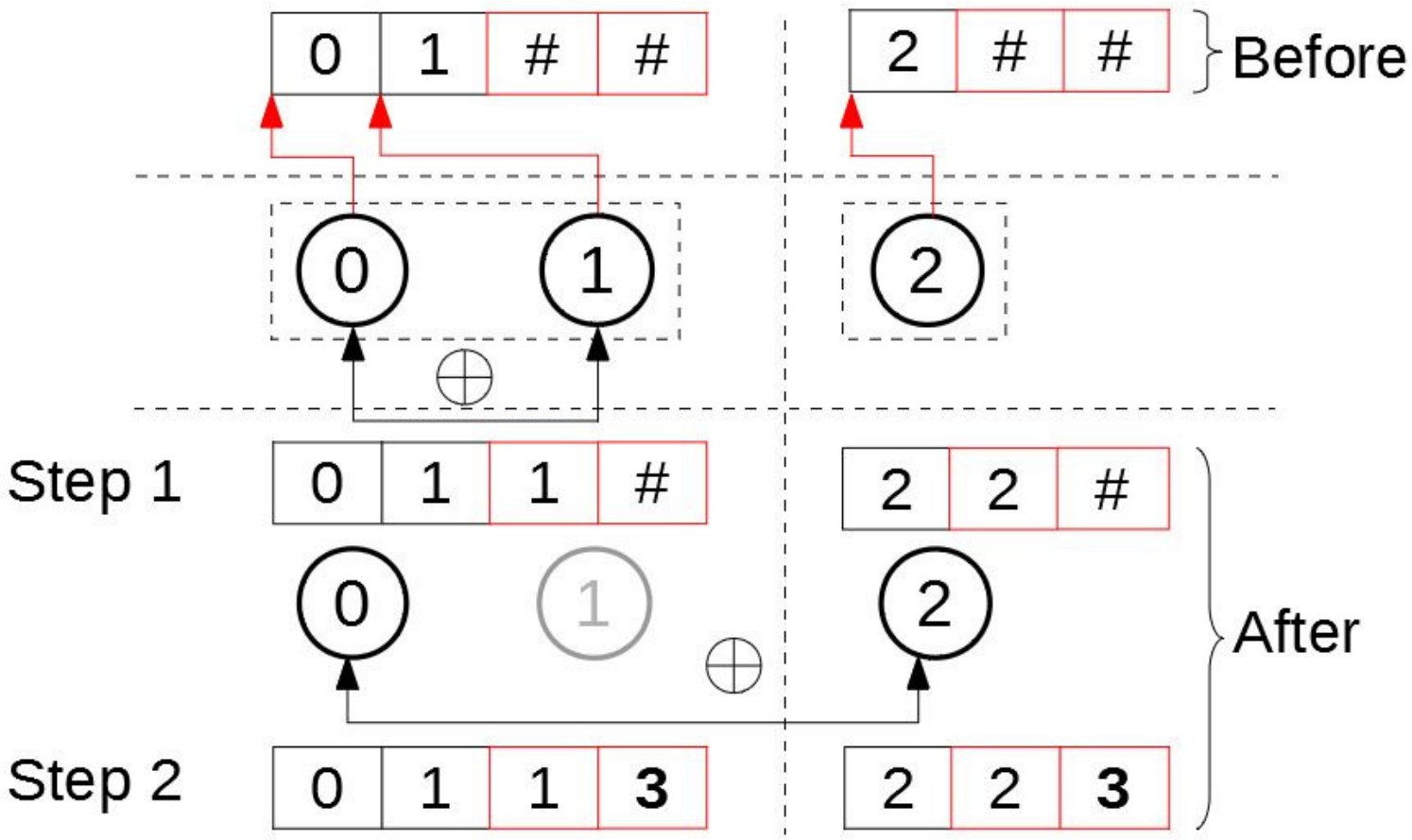}
\caption{Illustration of the hybrid MPI+MPI context-based \textit{allreduce}.
The input data is enclosed with black cubic
and the reduced results -- either locally or globally -- 
are enclosed with red cubic, 
where the globally reduced ones are stressed with bold font.
$\oplus$ means {\em MPI\_SUM} operation, which
is applied to the input and locally reduced data.
Refer to Figure~\ref{fig:allgathernativehy} for the explanations of arrows.}
\label{fig:allreducehy}
\end{figure}

The implementation method of the hybrid MPI+MPI context-based \textit{allreduce}
is illustrated in Figure~\ref{fig:allreducehy}.
Each process points to an element (as an input), which is located in
the shared region and supposed to be updated by its affiliated process.
Besides, an output vector with $2$ elements is appended to store the 
locally and globally reduced results, respectively.
Rather, the reduction computation in step $1$ proceeds at the node level.
Step $2$ is performed by all $leaders$ and 
applies the sum operation to the first elements with
the final result stored in the second element of the output vector.
This output vector is shared and can be accessed by all processes
on the same node, but only under a proper synchronization to secure
the computational determinacy.
The reduced result is thus not necessarily broadcast to 
all other on-node processes.
Clearly, the order of operands in our \textit{allreduce} approach is not defined to
be in ascending order of process rank beginning with zero.
In this example with block-style placement, 
we take advantage of the associativity of the 
sum operation to guarantee the correction of the reduced result.
However, the operation should be both
commutative and associative when a non-block-style placement is applied.
This \textit{allreduce} approach is thus valid for all predefined operations,
which are assumed to be commutative as well as associative.

\begin{figure}[h!]
\begin{lstlisting}[mathescape, escapechar=']
template<class myType>
void Wrapper_Hy_Allreduce(myType* start_addr,
					myType** result_addr, int sharedmem_rank,
					int msize, MPI_Datatype data_type, MPI_Op op, 
					struct comm_package* comm_handle, MPI_Win win);
\end{lstlisting}
\caption{The template interface to {\em Wrapper\_Hy\_Allreduce}.}
\label{fig:impl:allreduce}
\end{figure}
The template function {\em Wrapper\_Hy\_Allreduce} acts as the counterpart to the 
\textit{allreduce} in the pure MPI version.
Each node is required to contribute an intermediate result to the
output vector in step $1$, which can be completed in two ways.
One is letting the $leader$ serially perform the sum operation on 
an element-wise basis, which however leads to the issue of core idles
and extra synchronization (explained in Section~\ref{sec:impl-sync}).
The other is performing an {\em MPI\_Reduce} to return the \mbox{locally-reduced}
result to the $leader$, which implies a synchronization point among on-node processes,
and however brings MPI internal memory copies.
Step $2$ proceeds with a standard \textit{allreduce} operation called by all $leaders$.
Figure~\ref{fig:impl:allreduce} shows the template interface to {\em Wrapper\_Hy\_Allreduce}, where
the input parameters {\em sharedmem\_rank} and {\em win} are responsible
for the identification of local pointer in step $1$ and 
the synchronization operations after step $2$, respectively.

\subsection{Synchronization consideration}
\label{sec:impl-sync}

\newsavebox{\allgathersync}
\begin{lrbox}{\allgathersync}
\begin{lstlisting}[mathescape, escapechar=']  
if (comm_handle->bridge_comm != MPI_COMM_NULL){// Leaders
			'\textcolor{red}{sync(comm\_handle->shmem\_comm)}';
			MPI_Allgatherv(sbuf,rbuf,...,comm_handle->bridge_comm);
			'\textcolor{green!10!orange!90!}{sync(comm\_handle->shmem\_comm)}';}
else{// Children
			'\textcolor{red}{sync(comm\_handle->shmem\_comm)}';
			'\textcolor{green!10!orange!90!}{sync(comm\_handle->shmem\_comm)}';}
\end{lstlisting}
\end{lrbox}

\newsavebox{\bcastsync}
\begin{lrbox}{\bcastsync}
\begin{lstlisting}[mathescape, escapechar=']
if (comm_handle->bridge_comm != MPI_COMM_NULL){// Leaders
			MPI_Bcast(buf, ..., comm_handle->bridge_comm);
			'\textcolor{green!10!orange!90!}{sync(comm\_handle->shmem\_comm)}';}
else// Children
			'\textcolor{green!10!orange!90!}{sync(comm\_handle->shmem\_comm)}';
\end{lstlisting}
\end{lrbox}

\newsavebox{\allreducesync}
\begin{lrbox}{\allreducesync}
\begin{lstlisting}[mathescape, escapechar=']
/* Step 1 */
Method 1:
MPI_Reduce(..., comm_handl->shmem_comm);
Method 2:
'\textcolor{red}{sync(comm\_handle->shmem\_comm)}';
Each leader applies the operation on node-level;
/* Step 2 */
if (comm_handle->bridge_comm != MPI_COMM_NULL)
			MPI_Allreduce(sbuf,rbuf,..., comm_handle->bridge_comm);
'\textcolor{green!10!orange!90!}{sync(comm\_handle->shmem\_comm)}';
\end{lstlisting}
\end{lrbox}

\begin{figure}[h!]

\subfloat[\textit{allgather}]{\label{sync:allgather}\usebox{\allgathersync}} \\
\subfloat[broadcast]{\label{sync:bcast}\usebox{\bcastsync}} \\
\subfloat[\textit{allreduce}]{\label{sync:allreduce}\usebox{\allreducesync}}

\caption{Three pieces of pseudo-code handling with the synchronization among on-node processes
for our \textit{allgather}, broadcast and \textit{allreduce}.}
\label{fig:sync}
\end{figure}

\begin{figure}[h!]
\noindent\begin{minipage}{.22\textwidth}
\begin{lstlisting}[numbers=none, escapechar=']
Leader:
			'$status$' = 0;
			compute&communication;
			'$status$'++;
			MPI_Win_sync(win);
\end{lstlisting}
\end{minipage}
\noindent\begin{minipage}{.24\textwidth}
\begin{lstlisting}[numbers=none]
Children:		
			ref = 0; ref++;
			while(1){
					MPI_Win_sync(win);
					if(status==ref) break;}
\end{lstlisting}
\end{minipage}

\caption{Pseudo-code that demonstrates the spinning method.}
\label{fig:spinningcode}
\end{figure}

The synchronization and communication among processes are more
decoupled in hybrid MPI+MPI, than those in pure MPI.
Therefore, the synchronization operations need to be explicitly added to
guarantee the data integrity and support a deterministic computation
in hybrid MPI+MPI.
This section supplements the illustration of our collectives
with due consideration of the node-level synchronization points,
which are intuitively marked with $sync$ in Figure~\ref{fig:sync}.
The calls to $sync$ are highlighted
using two kinds of colors featuring different synchronous patterns.

Next, we shed lights on the two different synchronous patterns reflected
in the above three MPI+MPI context-based collective functions (prefixed with \texttt{Wrapper\_Hy}),
which are assumed to execute on more than one node.
We start with the implementation of the function {\em Wrapper\_Hy\_Allgather},
where two $sync$ calls among all the on-node processes need to 
be added before and after the irregular \textit{allgather} operation, respectively.
The first $sync$, shown in red, guarantees that all processes finish 
the updates to the shared data that has affinity to them.
The second $sync$, shown in yellow, is invoked to block the $children$ until
the $leaders$ exit from the irregular \textit{allgather} operation.
Figure~\ref{sync:allgather} shows its implementation.
Then it comes to the function {\em Wrapper\_Hy\_Bcast}, in which a $sync$ operation
is needed after the broadcast operation to guarantee that the broadcast 
data is ready for all the on-node processes. Figure~\ref{sync:bcast} shows the related pseudo-code.
This is followed by the function {\em Wrapper\_Hy\_Allreduce}, characterizing
the two methods of the intermediate reduction among on-node processes.
The $method$ $1$ is adopted to return the reduced result to each $leader$
for simplicity and flexibility, which could bring performance issues due to
the MPI internal buffering policy.
Instead of calling {\em MPI\_Reduce}, we 
can use an ad hoc method ($method$ $2$).
It adds a $sync$ operation to guarantee that all the input data is ready to be used by
the $leader$ to compute the reduced result.
In addition, the second $sync$ comes to let the $children$ wait for 
the completion of the \textit{allreduce}
operation called by $leaders$. 
Its implementation is shown in Figure~\ref{sync:allreduce}.

Based on the Figure~\ref{fig:sync}, we can draw a general conclusion that 
the $sync$ in red entails a collective synchronization among a set of processes
and the $sync$ in yellow can be treated as a lightweight one in comparison to the former.
Rather, with the $sync$ in red, each process must stop at this point
until all other processes reach this $sync$. The function {\em MPI\_Barrier}
is thus applicable to this $sync$. And yet
all the $children$ must pause until their $leader$
reaches the $sync$ in yellow.
In short, the $sync$ in yellow synchronizes the $leader$ with its $children$.
If this $sync$ is also substituted by a barrier, 
the $children$ will end up waiting for each other, which implies
unnecessary handshaking and leads to severe degradation of performance.
This $sync$ occurs after a barrier point in terms of our \textit{allgather} and \textit{allreduce} approaches,
where the expected wait time for the $leaders$ should not be long.
Because in this situation the process skew is caused by the fact that only 
the $leaders$ participate in the collective operation.
Therefore, spinning in a loop~\cite{xiong2010ad} could be a simple as well as a more efficient
alternative synchronized method to the barrier.

To implement the spinning method, a shared variable (named $status$) is defined,
which can only be updated by the $leader$. The $children$ check the shared variable
by spinning in a polling loop. In other words, 
The $children$ do not exit the loop until the update to the shared variable meets an exit condition.
This spinning method is worthwhile only if the update to the shared variable
takes low clock cycles, otherwise performance issues will be caused, since
many cores waste time doing useless computation.
We thus simply use the increment ($++$) operator to modify the shared variable.
MPI establishes a restriction~\cite{MPISpec} on the concurrent access to the same shared memory location as
it does not support atomic  operators (such as increment) on numeric values requiring more than 
one byte.
This restriction permits polling on a shared memory location for a change from one
value to another value rather than comparing them.
In our implementation, the shared variable is contained within an MPI shared memory window.
Hence, the above restriction must be considered to guarantee a definite outcome and prevent the $children$ from being stuck in an endless loop.
The exit condition is then expressed as \enquote*{the shared variable == a certain value},
rather than as \enquote*{the shared variable $\geq$/$\leq$ a certain value}.
Note that the routine {\em MPI\_Win\_sync}  must be included by
both the $leader$ and its $children$ to achieve a processor-memory barrier (see Section~\ref{sec:background:sharedmem}).

\section{Evaluation}
\label{sec:eval}
In this section, we compare the performance characteristics of 
the hybrid MPI+MPI programs (including our collectives)
with the pure MPI and hybrid MPI+OpenMP programs
(containing the standard MPI counterparts). 
Our studies were conducted on two parallel clusters by measuring the
latencies with a varying number of cores and different message sizes.
We first briefly describe our experimental testbed, then
discuss the overheads of the aforementioned one-off 
activities, and finally evaluate the performance of the micro-benchmarks
and application kernels.
These micro-benchmarks were developed according to the 
OSU benchmark\footnote{http://mvapich.cse.ohio-state.edu/benchmarks/}
and averaged over $10,000$ executions.
The kernel-level experiments
consist of a computation with
Scalable Universal
Matrix Multiplication Algorithm (SUMMA), 2D Poisson solver and Bayesian Probabilistic Matrix
Factorization (BPMF).
We used the default MPI rank placement scheme 
-- block-style -- to run all these benchmarks.

\subsection{Experimental setup}
\label{sec:eval:setup}
We used a Cray XC40 and a NEC cluster for our experiments:
\begin{enumerate}
\item Cray XC40 (aka. Hazel Hen):
Each of the Hazel Hen compute nodes has $24$ Intel
Haswell cores running at \SI{2.5}{\giga\hertz} with
\SI{128}{\giga\byte} of DDR4 main memory.
The cores are organized as two sockets with $12$ cores per socket
(each socket is seen as a NUMA domain).
The nodes are connected with dedicated Cray Aries network which has
a dragonfly topology.
The GNU programming environment 6.0.5 and
the version of cray-mpich/7.7.6 were applied to this system.

\item NEC cluster (aka. Vulcan):
Vulcan consists of several compute nodes of different types.
We used SandyBridge (SB) and Haswell compute nodes.
Each of the SB compute node has in total
$16$ SB cores running at 
\SI{2.6}{\giga\hertz} with \SI{64}{\giga\byte} DDR3 main memory
($8$ cores per NUMA domain).
The configuration of the Haswell compute node is the same as above.
The applied GNU compiler version was 8.3.0.
The nodes are connected via the InfiniBand network. 
The version of Open MPI/4.0.1 was run.
\end{enumerate}

\subsection{Microbenchmark evaluation}
All the microbenchmark evaluations 
were executed on both of the two clusters and described in two aspects:
the overhead caused by our design (called implementation overhead below)
and the performance comparison
between the standard MPI collectives and their counterparts (our approaches)
in the hybrid MPI+MPI context.
For brevity
we mostly present the evaluation results on Vulcan with Haswell compute nodes.
But we will go into details when the results on Hazel Hen are different from those on Vulcan.
The labels prefixed with \texttt{Wrapper\_Hy} in the following figures indicate 
our collectives, otherwise they refer to the standard ones.

\subsubsection{Implementation overhead}
\label{sec:eval:overhead}

Table~\ref{tab:overhead} displays the implementation overhead imposed by our design on Vulcan with
the leftmost columns listing the primitives.
Besides the common primitives of two-level communicator splitting (\texttt{Communicator})
and shared memory allocation (\texttt{Allocate}),
the primitives of {\em Wrapper\_Get\_transtable} and {\em Wrapper\_Create\_Allgather\_param}
are abbreviated as the the \texttt{Bcast\_transtable} and \texttt{Allgather\_param}, respectively.
Their overheads are subject to the number of cores rather than message sizes.
Hence, their overheads over the number of cores ($16$, $64$, $256$ and $1,024$)
are given to investigate the scalability of these wrapper functions.

\begin{table}[thp]
\footnotesize
\begin{center}
\begin{tabular}{|c|c|c|c|c|c|}
\hline
\multicolumn{2}{|c|}{\multirow{3}{*}{Primitives}}                                       & \multicolumn{4}{c|}{\#Cores}                                                                                                                                                                                          \\ \cline{3-6} 
\multicolumn{2}{|c|}{}                                                                     & 16                                                  & 64                                                  & 256                                                 & 1024                                                \\ \cline{3-6} 
\multicolumn{2}{|c|}{}                                                                     & \begin{tabular}[c]{@{}c@{}}Mean\\ (us)\end{tabular} & \begin{tabular}[c]{@{}c@{}}Mean\\ (us)\end{tabular} & \begin{tabular}[c]{@{}c@{}}Mean\\ (us)\end{tabular} & \begin{tabular}[c]{@{}c@{}}Mean\\ (us)\end{tabular} \\ \hline
\multirow{2}{*}{Common} & \texttt{Communicator}                                                      & 64.8                                                & 170.9                                               & 413.7                                               & 1098.7                                              \\ \cline{2-6} 
                         & \begin{tabular}[c]{@{}c@{}}\texttt{Allocate} \end{tabular} & 188.3                                               & 262.5                                               & 307.1                                               & 311.8                                               \\ \hline
\multicolumn{2}{|c|}{\texttt{Bcast\_transtable}}                                                                & 0.7                                                 & 9.2                                                 & 95.9                                                & 1462.8                                              \\ \hline
\multicolumn{2}{|c|}{\texttt{Allgather\_param}}                                                            & 0.3                                                 & 2.9                                                 & 7.1                                                 & 19.9                                                \\ \hline
\end{tabular}
\end{center}

\caption{One-off overheads associated with the hybrid MPI+MPI programs containing collective operations.}
\label{tab:overhead}
\end{table}
The rows of  \texttt{Communicator} and
\texttt{Bcast\_transtable} show that their overheads increase nearly
proportionally to the number of cores.
The \texttt{Allocate} shows good scalability but its
overhead should still be analyzed in a hybrid MPI+MPI program.
The overhead in regard to our \textit{allgather} approach -- shown in
the last row -- is almost negligible.
The evaluation on Hazel Hen shew similar implementation overheads
as were shown on Vulcan, except the overheads for \texttt{Communicator}
and \texttt{Bcast\_transtable} were one magnitude fewer.
All of these overheads can be treated as one-offs, which means they will
not repeatedly be added up to the total elapsed time of a hybrid MPI+MPI program.
Nevertheless, we need to check the effectiveness of applying the hybrid 
MPI+MPI mechanism to an application
by analyzing the occurrence frequency and accumulated overhead of the collective operation.
We should thus guarantee that the implementation overheads are traded for
the greater performance benefits of our collectives' counterparts (i.e., {\em Wrapper\_Hy\_Allgather},
{\em Wrapper\_Hy\_Bcast} and {\em Wrapper\_Hy\_Allreduce}).

\subsubsection{Allgather comparison}
\label{sec:eval-allgather}

\begin{figure}[hbp]
\begin{center}
\includegraphics[width=0.5\textwidth,height=0.20\textheight]{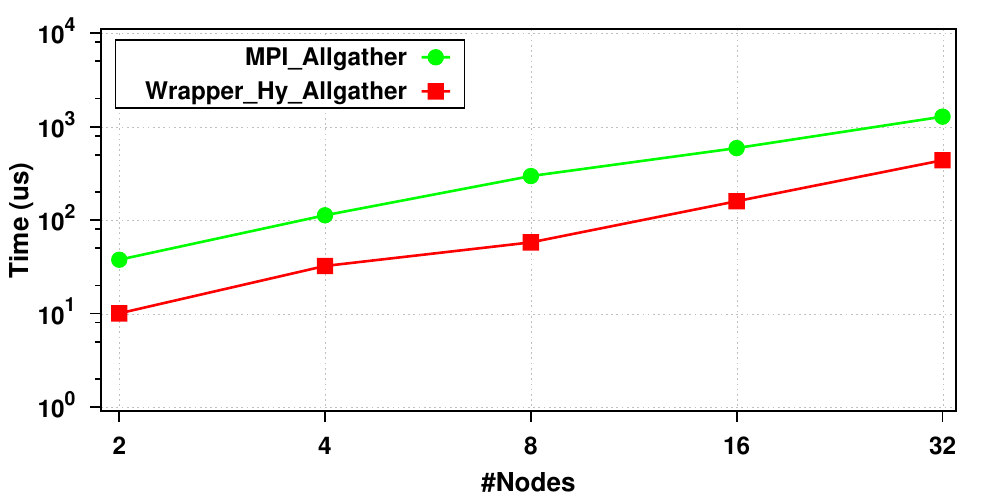}
\end{center}
\caption{The performance comparison between {\em Wrapper\_Hy\_Allgather} and {\em MPI\_Allgather}
on Hazel Hen with varying number of nodes. The size of the gathered message from every process is \SI{800}{\byte}.
}
\label{allgather:regular}
\end{figure}


Figure~\ref{allgather:regular} shows the time performance comparison
between {\em MPI\_Allgather} and {\em Wrapper\_Hy\_Allgather} 
on $2$, $4$, $8$, $16$ and $32$ nodes for a fixed
message length of \SI{800}{\byte}.
Here each of the nodes was populated with $24$ processes.
The same number of processes in different nodes leads to
a regular \textit{allgather} problem.
We can observe the advantage of our \textit{allgather} due to its constant
lower latencies. However,
the study of the performance characteristics of our proposed {\em Wrapper\_Hy\_Allgather}
does not merely discuss the regular problems.
The {\em MPI\_Allgatherv} suffers a performance penalty,
since its performance is determined by the maximum amount of data to
be received by a node~\cite{traff2009relationships}.
The irregular problem, where the number of MPI processes varies from node to node, 
is however a commonplace for our \textit{allgather} approach.
Hazel Hen is equipped with \mbox{non-power-of-two cores} ($24$) nodes throughout the system,
leading to irregularly-populated nodes when we request
power-of-two processes.
Our previous work~\cite{zhou2019mpi} confirms
a comprehensive insight into the performance benefits of our \textit{allgather}
approach for irregular as well as regular problem on Vulcan and Hazel Hen clusters.

\subsubsection{Broadcast comparison}
\label{sec:eval-bcast}

\begin{figure*}[h!]
\begin{center}
\includegraphics[width=\textwidth,height=0.17\textheight]{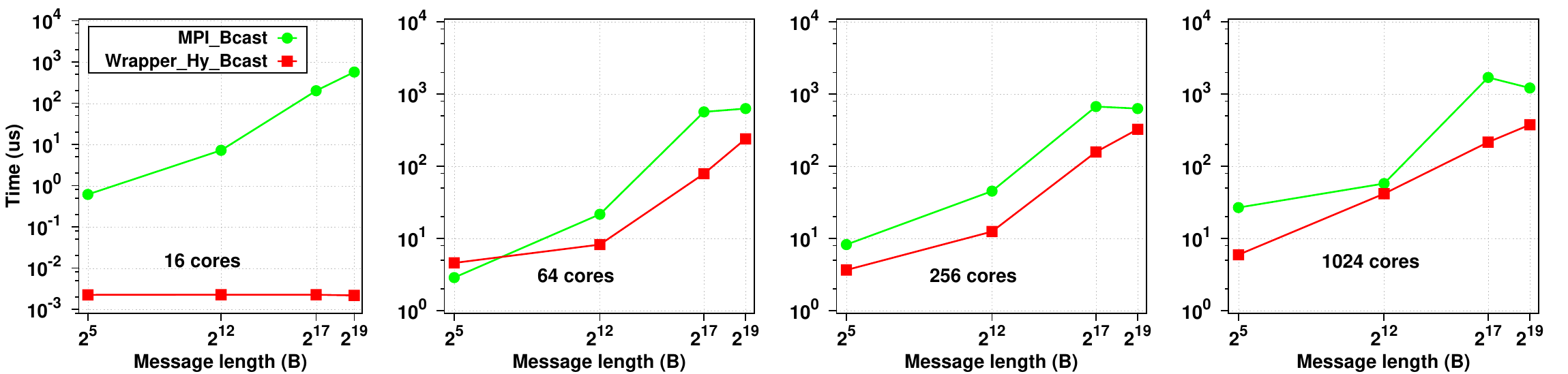}
\end{center}
\caption{The time performance comparison between {\em Wrapper\_Hy\_Bcast} and {\em MPI\_Bcast} on Vulcan
with varying numbers of cores and message lengths.}
\label{fig:bcastComm}
\end{figure*}

The Open MPI/4.0.1 supports several implementation algorithms for each of the
collective communication operations. The decision to switch between them depends
on the size of communicator as well as the message size.
More precisely, two message size thresholds, \SI{2}{\mykilobyte} and  $\sim$
\SI{362}{\mykilobyte}, are used in the Open MPI broadcast implementation.
We decided upon defining small, medium, and large message as 
$\leq \SI{2}{\mykilobyte}$, $> \SI{2}{\mykilobyte}$ and $\leq \SI{362}{\mykilobyte}$,
and $> \SI{362}{\mykilobyte}$
for the purpose of this experiment.

Figure~\ref{fig:bcastComm} compares the time performance of {\em MPI\_Bcast} and
{\em Wrapper\_Hy\_Bcast},
with $16$, $64$, $256$ and $1,024$ cores on Vulcan.
We varied the numbers of the broadcast elements of
double precision floating pointer (\SI{8}{\byte}) from $2^{0}$ to $2^{17}$ in this benchmark.
For brevity, we reported only the latency results for element counts of
$2^{2}$, $2^{9}$, $2^{14}$ and $2^{16}$, which represent small (\SI{32}{\byte}),
medium (\SI{4}{\mykilobyte} and \SI{128}{\mykilobyte}) and large messages (\SI{512}{\mykilobyte}), respectively.
The current version of {\em Wrapper\_Hy\_Bcast} replaces
the synchronization point with a barrier operation.
We observe that our proposed broadcast approach offers significantly
lower latency than the standard one, 
except for the small message running on $64$ cores.
This is probably because the synchronization overhead contributes
more to the latency of broadcast than the data transfer overhead.
The impact of the synchronization point on the performance
of our collectives will be discussed at length in Section~\ref{sec:eval-allreduce}.
On Hazel Hen, the standard broadcast was always inferior to our approach.
The first subplot shows the results running on $16$ cores,
where all the MPI processes reside on the same node and thus
no inter-node data exchanges will be involved.
In this scenario,
only an {\em MPI\_Barrier} is called by the on-node processes
and as we expected, 
its latency almost keeps constant, regardless of the message lengths.
The remaining three subplots show the latency results across different nodes,
wherein the curves go up steadily as the broadcast message size grows except when
the message size reaches \SI{512}{\mykilobyte}.
Such exception happens, since the broadcast algorithm is changed from \textit{split binary
tree}~\cite{pjesivac2007towards} to \textit{pipeline}.

\subsubsection{Allreduce comparison}
\label{sec:eval-allreduce}

\begin{figure*}[h!]
\begin{center}
\includegraphics[width=\textwidth,height=0.17\textheight]{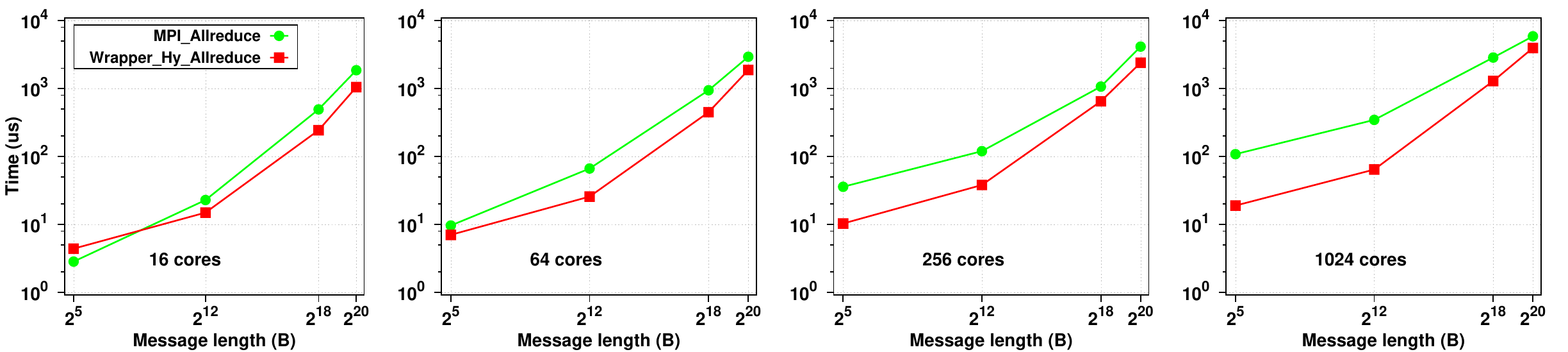}
\end{center}
\caption{The time performance comparison between {\em Wrapper\_Hy\_Allreduce} and {\em MPI\_Allreduce} on Vulcan with varying numbers of cores and message lengths.}
\label{fig:allreduceComm}
\end{figure*}

The intermediate message threshold ($\sim$\SI{9}{\mykilobyte}) defined
in the Open MPI \textit{allreduce} implementation roughly determines
whether the involved message in this experiment is small, medium or large.

Figure~\ref{fig:allreduceComm} compares the time performance of
{\em MPI\_Allreduce} and {\em Wrapper\_Hy\_Allreduce} on Vulcan, as either
the message length or the number of cores grows.
We ran this experiment with the increasing
number of elements of double precision floating point, where $2^{2}$, $2^{9}$, $2^{15}$ and $2^{17}$
were chosen as the representatives for the small (\SI{32}{\byte}),
medium (\SI{4}{\mykilobyte}) and large message (\SI{256}{\mykilobyte} and \SI{1}{\mega\byte}).
The version of {\em Wrapper\_Hy\_Allreduce} for Figure~\ref{fig:allreduceComm}
used \mbox{$method$ $1$} to implement step $1$ and replaced
the synchronization point in step $2$ with a barrier call.
Not surprisingly, our \textit{allreduce} approach fails to significantly
outperform the standard one for small messages on $16$ cores.
Otherwise, speedups (range from $27.2\%$ to $82.5\%$)
of our \textit{allreduce} over the standard approach can be
achieved anywhere.
On Hazel Hen, our \textit{allreduce} performed worse than the standard approach for
small-size messages up to \SI{2}{\mykilobyte} on all the above number of cores.
The inferiority of our \textit{allreduce} for small messages is attributed to
inadequate methods for step $1$ or inefficient synchronization implementation.
The data transfer latency for small messages is very low and thus strongly
affected by the overhead of the synchronization operation.
However, in an application the involved
\textit{allreduce} operation could be used with messages
centering on the sizes smaller than \SI{1}{\mykilobyte}~\cite{techrep/HYCOMperf}.
This drives us to evaluate the performance of the $method$ $2$ (for step $1$)
and the spinning method (for step $2$).
Using $method$ $2$ instead will slightly improve the performance of our approach on
both Vulcan and Hazel Hen only for a range of small message sizes.
Then again, adopting the spinning method can noticeably lessen the latency of
our \textit{allreduce}, especially for small messages, on both Vulcan and Hazel Hen.
For large messages, we observed that our \textit{allreduce} latencies of both  versions using
barrier and spinning are at the same level.
This is because, for large messages the synchronization time becomes insignificant
and then the data transfer overhead dominates the latency of our \textit{allreduce}.
Therefore, the current version of {\em Wrapper\_Hy\_Allreduce} replaces 
the synchronization point
in step $2$ with the spinning method and 
chooses between $method$ $1$ and $method$ $2$ 
in terms of message sizes for optimal performance.
\begin{figure}[h!]
\begin{center}
\subfloat[Vulcan]{\label{vulcanallreduceoptcomp:16}\includegraphics[width=0.48\textwidth,height=0.17\textheight]{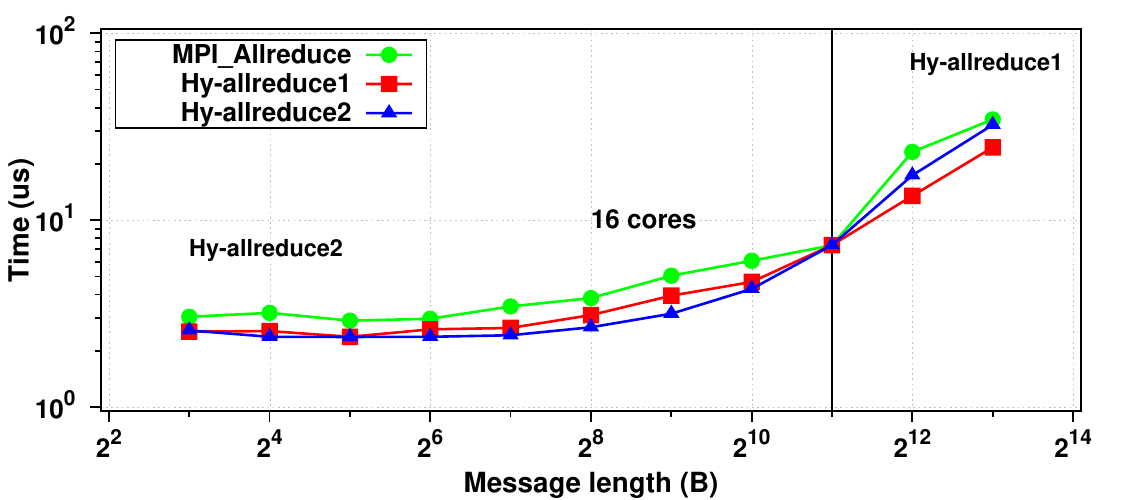}} \\
\subfloat[Hazel Hen]{\label{hazelhenallreduceoptcomp:16}\includegraphics[width=0.48\textwidth,height=0.17\textheight]{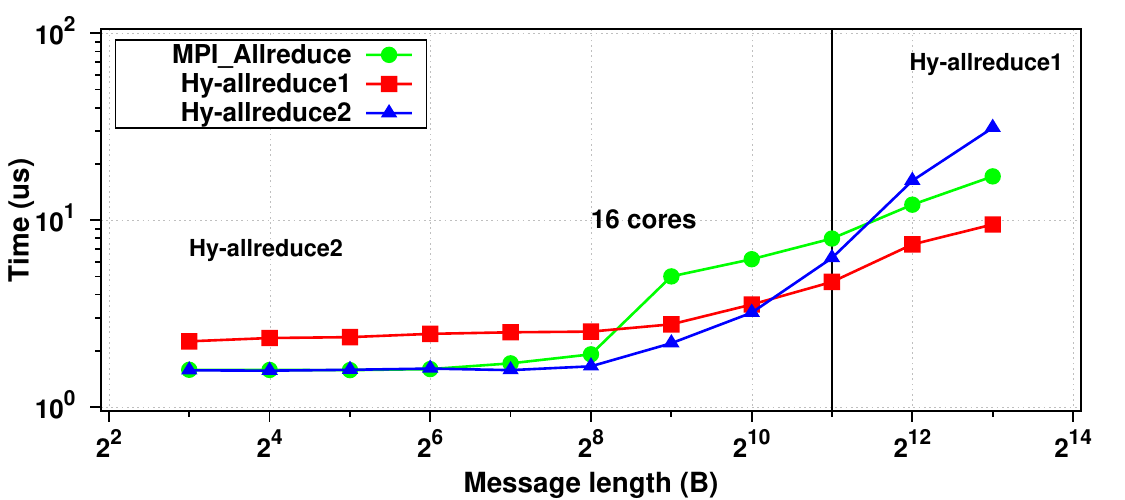}}

\end{center}
\caption{The time performance comparison of {\em Hy-allreduce1}, {\em Hy-allreduce2}
and {\em MPI\_Allreduce} for small messages on a single node ($16$ cores) on Vulcan and Hazel Hen.}

\label{allreduceoptcomp}
\end{figure}

\begin{figure}[h!]
\begin{center}
\includegraphics[width=0.4\textwidth,height=0.17\textheight]{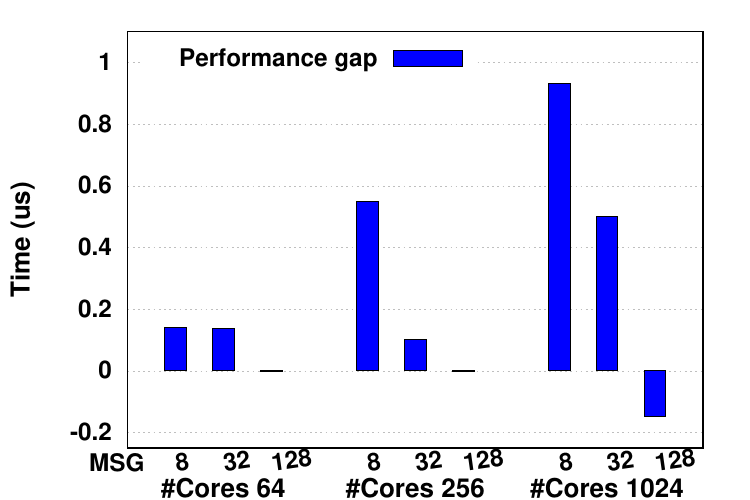}
\end{center}
\caption{The performance gap between {\em MPI\_Allreduce} 
and the optimized {\em Wrapper\_Hy\_Allreduce} on Hazel Hen.}
\label{allreduce:performancegap}
\end{figure}

Next, we develop two versions of our \textit{allreduce}, one with $method$ $1$
and the other with $method$ $2$, 
to determine the cut-off value of the message size for switching
from $method$ $2$ to $method$ $1$ in the optimal version of
our \textit{allreduce}.
Henceforth the versions with $method$ $1$ and $method$ $2$ are abbreviated as
{\em Hy-allreduce1} and \mbox{{\em Hy-allreduce2}}, respectively.
Figure~\ref{allreduceoptcomp} compares the latency of
{\em Hy-allreduce1} and {\em Hy-allreduce2} on $16$
cores, which all reside on the same node.
This experiment ran with the message sizes ranging from \SI{8}{\byte}
to \SI{8}{\mykilobyte} on Vulcan and Hazel Hen.
Besides, the performance curve for {\em MPI\_Allreduce} is added as baseline.
Obviously, the cut-off value of the message size is \SI{2}{\mykilobyte},
which is marked with a vertical line.
The {\em Hy-allreduce2} performs slightly better before the cut-off point 
and becomes worse when it is surpassed.
Therefore, our \textit{allreduce} is further optimized to use
$method$ $2$ and $method$ $1$ before and after the cut-off point, respectively.
Figure~\ref{fig:allreduceComm} already shows us the scalability
of the initial version of our \textit{allreduce} with the increasing number of cores on Vulcan.
It was therefore necessary to reevaluate the scalability after the above tuning.
The curve trends presented in the new plots (omitted for brevity), that we obtained during reevaluation, coincided
well with those displayed in Figure~\ref{fig:allreduceComm}, except for
the first subplot with $16$ cores.
Therefore, the results reflected in Figure~\ref{fig:allreduceComm} are also partially fit for
further reference in Section~\ref{sec:eval-poisson}.
We then compute the performance gap between our optimized \textit{allreduce} and 
the standard approach on Hazel Hen for $64$, $256$ and $1,024$ cores
respectively.
The results are shown in Figure~\ref{allreduce:performancegap},
where the label \textit{MSG} denotes the message length in bytes.
From this figure, we can observe that the standard \textit{allreduce} still slightly outperforms
our \textit{allreduce} for \SI{8}{\byte} and \SI{32}{\byte}.
The negative performance gap at \SI{128}{\byte} means that our \textit{allreduce}
starts to perform better than the standard approach.

\subsection{Kernel-level benchmarks}
\label{sec:eval-app}
In this section, we consider three benchmarks -- SUMMA, 2D Poisson solver and BPMF --
that have different collective communication operations interweaving with real computations.
The BPMF was executed on Hazel Hen with Haswell compute nodes (each contains $24$ cores)
while the SUMMA and 2D Poisson solver were run on Vulcan with SB compute nodes, of which
each contained power-of-two ($16$) cores.
Each node was fully populated with MPI processes or OMP threads when the three
benchmarks were executed below.
The experimental results for SUMMA and 2D Poisson solver were
the average of at least $20$ runs and those for BPMF were the average of $3$ runs.
All of them shew standard deviations of only a few percentages.
For each benchmark we assessed the time performance and LOC of the pure MPI, hybrid
MPI+OpenMP and hybrid MPI+MPI implementations, where the former two 
utilized the standard MPI primitives to implement the relevant collective operations
and the last one utilized our wrapper primitives.
Specifically, we simply used the loop-level parallelization in the 
hybrid MPI+OpenMP implementations without putting great efforts into
achieving optimal performance.
It is to be noted that we only launched one MPI process per node and then 
this MPI process spawned threads fully populating the available
core resources on each node when running them, regardless of whether they were run on Vulcan or Hazel Hen. 
A thread was pinned to a specific core and this pinning went
successively through available cores.
To achieve this, the environment variables \texttt{OMP\_PLACES} and \texttt{OMP\_PROC\_BIND}, and
the option of \texttt{--map-by} needed to be correctly set on Vulcan and 
the \texttt{aprun} option of \texttt{-d} was given to specify the number of threads per MPI process on Hazel Hen.

The \textit{total} time shown in Figures \ref{fig:summa_vulcan} to \ref{fig:bpmf_hazelhen} below
is described as the sum of computation overhead
and relevant collective communication latency.
This can facilitate us to intuitively comprehend the impacts of the latter
on the \textit{total} performance and scalability of our benchmarks.
The \textit{total} here denotes the core part including intensive computation
and collective communication operations in each benchmark.
 
\subsubsection{SUMMA}
\label{sec:eval-SUMMA}

SUMMA multiplies two dense matrices by using a scalable universal algorithm
\cite{journals/concurrency/GeijnW97}.
In this kernel, two square matrices of the same type (double-precision) and size
are required as input data
and evenly decomposed into blocks, each of which is assigned to an MPI process.
This kernel is a typical example of supporting multiple communicators in 
our design. Herein
we first logically laid out the {\em MPI\_COMM\_WORLD} into a two-dimensional Cartesian grid
and then created sub-communicators for rows and columns.
This kernel consists of multiple core phases, whose elapsed time is
our measurement target.
In each core phase, two broadcast operations on the row and column sub-communicators 
are triggered due to the dependencies on the blocks living on the other MPI processes.

\begin{figure}[tbp]
\begin{center}
\includegraphics[width=0.48\textwidth,height=0.24\textheight]{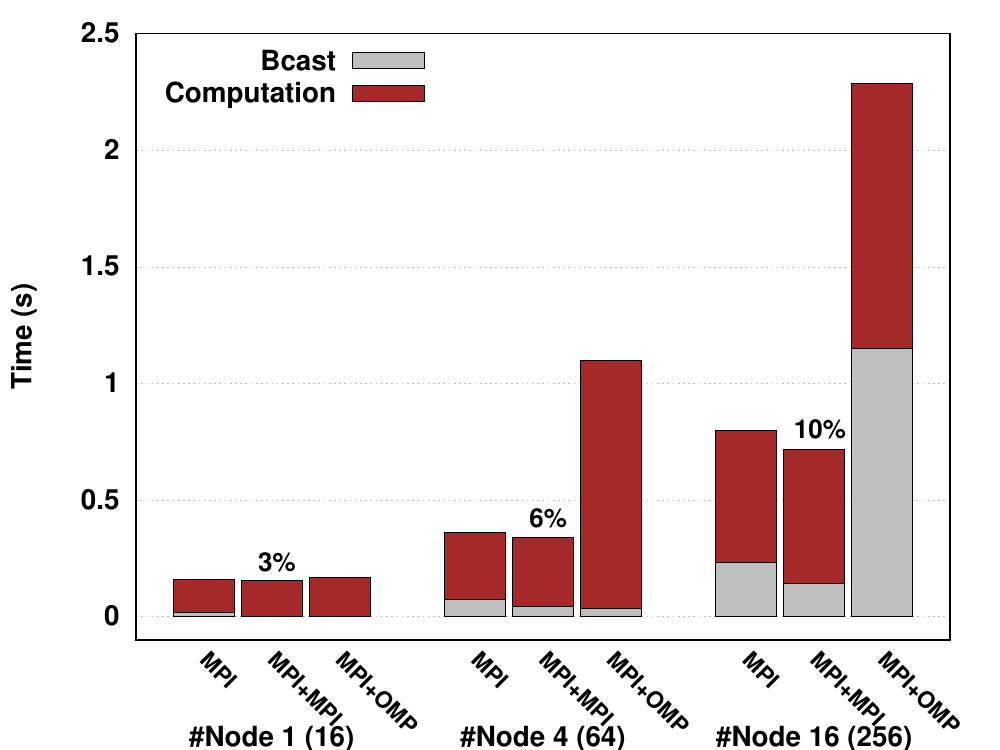}
\end{center}
\caption{The time performance comparison between different implementations of SUMMA on Vulcan.}
\label{fig:summa_vulcan}
\end{figure}

We ran all the three implementations
(pure MPI, MPI+MPI, and MPI+OpenMP) on Vulcan using
three matrices of size $1,024 \times 1,024$, 
$2,048 \times 2,048$ and $4,096 \times 4,096$, each with
$1$, $4$ and $16$ nodes, respectively.
The corresponding number of cores are indicated in parenthesis,
shown in Figure~\ref{fig:summa_vulcan}.
The same is true of Figures \ref{fig:2dPoisson_vulcan} and \ref{fig:bpmf_hazelhen}.
Figure~\ref{fig:summa_vulcan}
demonstrates the elapsed time of
the core phases of the three SUMMA implementations, where the
broadcast message size is \SI{512}{\mykilobyte}.
The comparison results on Hazel Hen can be found in \cite{zhou2019mpi}.
We observe that the hybrid MPI+OpenMP implementation indeed brings the minimal
broadcast overhead, but its computation overhead is greater than the other two implementations.
However, the broadcast message size in the hybrid MPI+OpenMP implementation is
always larger than those in the other two implementations due to the fewer number of MPI processes.
This disparity can lead to an exception -- the hybrid MPI+OpenMP implementation
delivers larger broadcast overhead than the other two -- on 16 nodes.
More significantly, the hybrid MPI+MPI implementation consistently has the best 
performance of the three SUMMA implementations.
Further, the improvements (i.e., $3\%$, $6\%$, and $10\%$) of the hybrid MPI+MPI implementation
over the pure MPI one are explicitly given in this figure.
This is not unexpected, since the hybrid MPI+MPI implementation 
constantly delivers less broadcast overhead in terms of 
the lower height of \textit{Bcast} bar. 
After revisiting the Figure~\ref{fig:bcastComm},
it can be found that our broadcast ({\em Wrapper\_Hy\_Bcast})
outperforms at \SI{512}{\mykilobyte}, from which 
we can infer that the superiority of this hybrid
MPI+MPI implementation over the pure MPI one is due to the usage of our broadcast method.
Compared with LOC for the pure MPI implementation,
the hybrid MPI+MPI implementation brings $6$ additional LOC for
an increase of $2\%$ in program size.

\subsubsection{2D Poisson solver}
\label{sec:eval-poisson}
\begin{figure}[tbp]
\begin{center}
\includegraphics[width=0.48\textwidth,height=0.24\textheight]{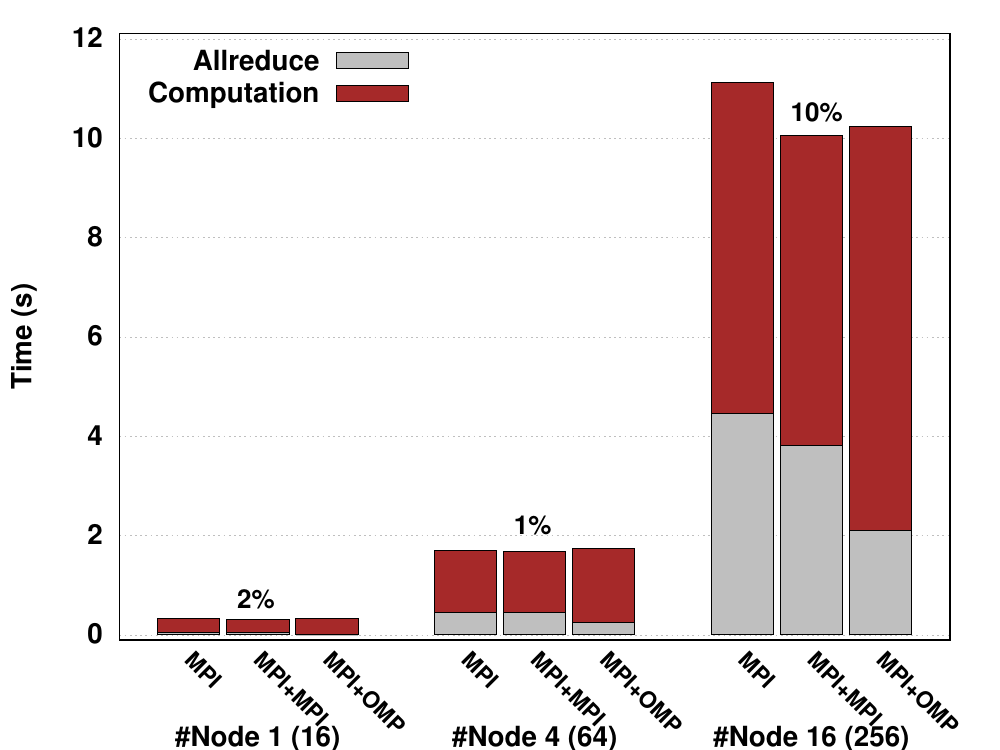}
\end{center}
\caption{The time performance comparison between different implementations of 2D Poisson solver
 on Vulcan.}
\label{fig:2dPoisson_vulcan}
\end{figure}

This kernel solves the 2D Poisson equation in an iterative way. A square grid 
holding elements of floating point is
initialized and evenly decomposed by rows among the MPI processes. 
In an iteration each MPI process first uses the Gauss-Seidel method
to do a five-point stencil computation on the current grid,
and then locally computes the maximum 
difference between the updated and exact grid,
and finally collectively calls the \textit{allreduce} operation to 
obtain the global maximum difference among all MPI processes.
This iteration is repeated until the global maximum difference is less than
a predefined convergence value. In this experiment,
the Gauss-Seidel module contains data transfers between adjacent processes
as well as the five-point stencil computation.
The data transfers are performed using a pair of MPI point-to-point
routines (i.e, {\em MPI\_Send} and {\em MPI\_Recv}).
The computation proceeds in the form of two nested loops.
We started our timing at the beginning of iterations
and stopped it until the convergence is reached.
We used three input grids of size
$256 \times 256$, $512 \times 512$ and $1,024 \times 1,024$, each running
on the number of nodes -- $1$, $4$ and $16$, respectively.
In Figure~\ref{fig:2dPoisson_vulcan}, we discuss the performance of
the 2D Poisson solver kernel.
The involved \textit{allreduce} operation is always used with small message
of \SI{8}\byte (aka. global maximum difference),
regardless of the grid size or node counts.
The curves in Figures~\ref{fig:allreduceComm} and \ref{vulcanallreduceoptcomp:16}
reveal that the performance benefits of our \textit{allreduce} over the
standard approach
are marginal on the small system (i.e., smaller than $64$ cores) and 
increase as the system size grows, for small messages (i.e.,
smaller than \SI{32}\byte).
This, in turn, explains that on $16$ nodes the hybrid MPI+MPI implementation
yields a $10\%$ time performance improvement over the pure MPI one, 
while it brings smaller performance gains of $2\%$ and $1\%$
on $1$ and $4$ nodes, respectively. We can also learn that
these performance advantages offered by the
hybrid MPI+MPI implementation are credited to
the application of our \textit{allreduce}.
The hybrid MPI+MPI implementation adds $7$ more additional LOC for
a code size increase of $1.6\%$, by comparison to the pure MPI one.

\subsubsection{BPMF}
\label{sec:eval-BPMF}

\begin{figure}[tbp]
\begin{center}
\includegraphics[width=0.48\textwidth,height=0.24\textheight]{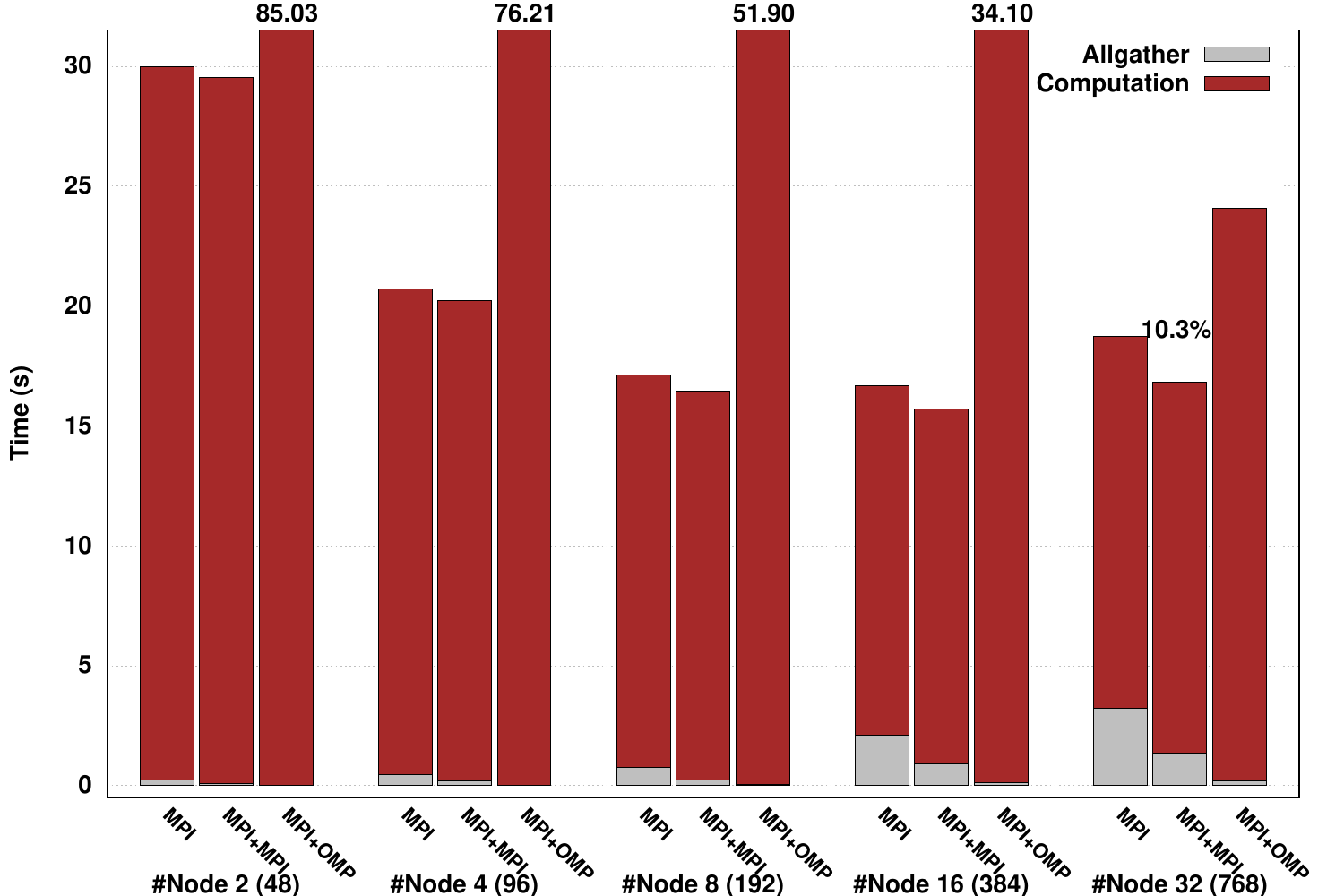}
\end{center}
\caption{The time performance comparison between different implementations of BPMF on Hazel Hen.}
\label{fig:bpmf_hazelhen}
\end{figure}

The BPMF kernel~\cite{ruslan:icml08, conf/cluster/AaCH16}
predicts compound-on-target activity
in chemogenomics based on machine learning.
The number of iterations to be sampled was
set to be $20$ for this experiment. Each iteration consists of two distinct
sampling regions on compounds and on-target activities followed by
a prediction. 
Both regions end with three calls to the regular \textit{allgather} operation.
In the three \textit{allgather} operations,
the sizes of the gathered messages from every process 
are \SI{80000}{\byte}, \SI{800}{\byte} and \SI{8}{\byte}, respectively.
The link\footnote{https://github.com/ExaScience/bpmf/} provides more information
about the BPMF code.
The strong scaling performance of the BPMF kernel with three different
implementations is demonstrated in Figure~\ref{fig:bpmf_hazelhen}. 
Here, the elapsed time of the sampled $20$ iterations is evaluated.
We used the $chembl\_20$ as our input training dataset, which is
a sparse matrix converted from ChEMBL publicly available data.
This kernel was run on different numbers of nodes, as shown in Figure~\ref{fig:bpmf_hazelhen}.
It is obviously observed that the hybrid MPI+MPI implementation is constantly
superior to the other two implementations.
The hybrid MPI+OpenMP implementation fails to be comparable to the other two, although
the performance gap between them shrinks as the system size increases.
The performance of both the pure MPI and hybrid MPI+MPI implementation degrades
when the node count increases from $16$ to $32$, since the increased
\textit{allgather} overhead overrides the decreased computation time.
Still, it is true that the performance of the pure MPI implementation
deteriorates more and the improvement of the hybrid MPI+MPI implementation over the pure MPI one
increases to $10.3\%$ on $32$ nodes. 
Figure~\ref{allgather:regular} implies that 
the application of our \textit{allgather} can take credit for the performance
advantage of the hybrid MPI+MPI implementation over the pure MPI one.
The hybrid MPI+MPI implementation brings $12$ extra LOC with an increase
of less than $0.1\%$ in code size, compared with the pure MPI one.

\section{Discussion and conclusion}
\label{sec:conclusion}

This paper proposes an innovative design method of
the collective communication operations (such as broadcast, \textit{Allgather}
and \textit{Allreduce}) that adapts to the hybrid MPI+MPI context,
and then describes them by assuming the block-style MPI rank placement.
With the other MPI rank placement schemes, 
our previous work~\cite{zhou2019mpi} discusses the measures that can be taken to
ensure the validity of our method.
Unlike the standard MPI collectives,
our collectives only maintain one copy of replicated data shared by
all on-node processes. In this way,
the explicit on-node inter-process data transfers are completely eliminated.
However, synchronization calls need to be adequately added
to guarantee the determinacy of the shared data among the on-node processes.
The micro-benchmark evaluations
present the overheads imposed by our implementation and
reveal that our collectives are
on par with or outperform those in the pure MPI context on Hazel Hen and Vulcan.
The synchronization overhead and its influence on the time performance
of \textit{allreduce} are also analyzed.
The application kernel evaluations show the superiority of the
hybrid MPI+MPI implementations to the other two implementations -- pure MPI and
hybrid MPI+OpenMP -- in time performance, which is credited to our collectives.
Further, the productivity gained from the hybrid MPI+MPI model can be comparable
to that gained from the pure MPI model in terms of LOC, owing to
the wrapper primitives that encapsulate all the implementation
details of our design from programmers.

For the evaluation results illustrated in Section~\ref{sec:eval-app}, further explanations are given.
First, the speedups of the hybrid MPI+MPI implementations are clearly quantified,
from which we observe that they are insignificant on a smaller number of nodes.
This does not necessarily mean that the performance advantage of our approach
increases as the system grows but instead the proportion of 
time spent in the collectives is greater.
Second, note that the obtained performance gains are kept above
the overall implementation overhead (see Table~\ref{tab:overhead}) in all the 
three hybrid MPI+MPI implementations.
Otherwise, the performance benefits caused by our collectives 
are pointless.

Our design method lacks in the distinction between the intra- and inter-NUMA data accesses,
since all the $children$ -- no matter they and their $leader$ are located in the same NUMA domain or not --
are limited to access the shared data allocated in the $leader$'s memory space.
To enable NUMA awareness, the most intuitive solution is to elect a $leader$ in each NUMA domain at the price
of maintaining a copy of replicated data in it. This comes with extra memory copes and thus needs to be
further investigation.
Note that our collectives may not apply to the applications using the master/slave pattern, where
the master needs to broadcast/gather a large amount of data to/from its slaves.
This is due to that the allocation of the MPI shared memory window fails when
the total amount of shared memory required by on-node processes exceeds a limit of size determined by
MPI implementation.
E.g., with Open MPI, the available shared memory space on a Haswell compute node (see Section~\ref{sec:eval:setup}) is
in the order of \SI{63}{\giga\byte}. 
Therefore, the use of our collectives is limited, but to a lesser extent.



\section{Acknowledgments}

The authors would like to thank 
Eric Gedenk for proofreading the article and
Tom Vander Aa for offering
the pure MPI implementation of the BPMF benchmark.
The comments made by the editor and reviewers are deeply appreciated.
Part of this work was supported by the European Community's Horizon 2020 POP project
[grant numbers~676553,~824080].








\bibliographystyle{elsarticle-num}
\bibliography{./paper}







\end{document}